\documentclass[12pt]{article}
\usepackage{amsmath,amssymb,leftidx,cite,epsfig,float}
\begin{document}

\title{\begin{flushright}
\small IISc/CHEP/02/12
\end{flushright}
\vspace{0.25cm} Renormalization of Noncommutative Quantum Field Theories}
\author{ 
Amilcar R. de Queiroz$^a$\footnote {amilcarq@unb.br}, Rahul Srivastava$^b$\footnote {rahul@cts.iisc.ernet.in} \, and Sachindeo Vaidya$^b$\footnote {vaidya@cts.iisc.ernet.in}
\\ 
\\
$^a$ \begin{small}
 Instituto de Fisica, Universidade de Brasilia, 
\end{small}\\
\begin{small}
Caixa Postal 04455, 70919-970, Brasilia, DF, Brazil
\end{small}
 \\
\\
$^b$ \begin{small}
 Centre For High Energy Physics, Indian Institute of Science, 
\end{small}\\
\begin{small}
Bangalore, 560012, India
\end{small}
 \\
}

\date{\empty}
\maketitle

\begin{abstract}
We report on a comprehensive analysis of the renormalization of noncommutative $\phi^4$ scalar field theories on the Groenewold-Moyal (GM) plane.  These scalar field theories 
are twisted Poincar\'e invariant. Our main results are that these scalar field theories are renormalizable, free of UV/IR mixing, possess the same fixed points and $\beta$-functions for
the couplings as their commutative counterparts. We also argue that similar results hold true for any generic noncommutative field theory with polynomial interactions and involving only pure matter fields. 
A secondary aim of this work is to provide a comprehensive review of different approaches for the computation of the noncommutative $S$-matrix:  noncommutative interaction picture and
 noncommutative LSZ formalism.

\end{abstract}

\vspace{2mm}

\section{Introduction}

Intuitive arguments involving standard quantum mechanics uncertainty relations suggest that at length scales close to Plank length strong gravity effects limit the spatial as well 
as temporal resolution beyond some fundamental length scale ($l_p$ $\approx$ Planck length), leading to space - space as well as space - time uncertainties \cite{dop}. One cannot 
probe spacetime with a resolution below this scale. That means that spacetime becomes fuzzy below this scale, resulting into noncommutative spacetime. Hence it 
becomes important and interesting to study in detail the structure of such a noncommutative spacetime and the properties of quantum fields written on it. It not only helps us 
improve our understanding of the Planck scale physics but also helps in bridging standard particle physics with physics at Planck scale.

There are various approaches to model the noncommutative structure of spacetime. The simplest one has coordinates satisfying commutation relations of the form
\begin{eqnarray} 
[\hat{x}_{\mu},\hat{x}_{\nu}] = i \theta_{\mu\nu} \; ; \qquad  \mu, \nu \, = \, 0,1,2,3,  \qquad \theta \: \; \text{a real, constant, antisymmetric matrix.}
\label{gm}
\end{eqnarray} 
The elements of the $\theta$ matrix have the dimension of $\text{(length)}^2$ and set the scale for the area of the smallest possible localization in the 
$\mu - \nu$ plane, giving a measure for the strength of noncommutativity \cite{dop1}. The algebra generated by $\hat{x}_{\mu}$ is usually referred to as the Groenewold-Moyal (GM) 
plane \cite{balreview}. In this paper we restrict ourself to the discussion of this noncommutative spacetime. Equivalently this noncommutative nature of spacetime can be taken 
into account by defining a new type of multiplication rule ($\ast$-product) between functions evaluated at the same point: 
 \begin{eqnarray} 
 f(x) \ast g(x) = f(x) e^{\frac{i}{2}\overleftarrow{\partial}_{\mu} \theta^{\mu\nu} \overrightarrow{\partial}_{\nu} } g(x).
\label{starproduct}
\end{eqnarray} 
One particularly important feature of GM plane, which makes it quite suitable for writing quantum field theories on it, is the restoration of Poincar\'e-Hopf symmetry as Hopf 
algebraic symmetry, by defining a new coproduct (twisted coproduct) for the action of the Poincar\'e group elements on state vectors \cite{wess,cha,cha1}.

Twisting of the coproduct has immediate implications for the symmetries of multi-particle wave functions describing identical particles \cite{sachin}. For example, on GM plane
 the correct physical two-particle wave functions are
\begin{eqnarray} 
  \phi \otimes_{S_{\theta}} \psi & \equiv &  \left( \frac{1 \, + \, \tau_{\theta}}{2} \right) ( \phi \otimes \psi ), \nonumber \\
\phi \otimes_{A_{\theta}} \psi & \equiv &  \left( \frac{1 \, - \, \tau_{\theta}}{2} \right) ( \phi \otimes \psi ),
\label{tsym}
 \end{eqnarray}
where $\phi$ and $\psi$ are single particle wavefunctions of two identical particles and $\tau_{\theta}$ is the twisted statistics (flip) operator associated with exchange of 
particles, 
given by 
\begin{eqnarray}
 \tau_{\theta} & = & \mathcal{F}^{-1} \tau_0 \mathcal{F}, \qquad \qquad \mathcal{F} \; = \; e^{\frac{i}{2} \theta^{\mu\nu} \partial_{\mu} \otimes \partial_{\nu} } .
 \label{tflip}
\end{eqnarray}
Here $\tau_0$ is the commutative flip operator : $\tau_0 \, ( \phi \otimes \psi ) \, = \,  \psi \otimes \phi$. 

The above analysis can be extended to field theories on GM plane resulting in a twist of the commutation relations between creation/annihilation operators \cite{sachin} :
\begin{eqnarray}
  a_{p_{1}} a_{p_{2}} & = & \eta \, e^{ip_{1}\wedge p_{2}} \,a_{p_{2}} a_{p_{1}}, \nonumber \\
a^{\dagger}_{p_{1}} a^{\dagger}_{p_{2}} & = & \eta \, e^{ip_{1}\wedge p_{2}} \, a^{\dagger}_{p_{2}} a^{\dagger}_{p_{1}}, \nonumber \\
a_{p_{1}} a^{\dagger}_{p_{2}} & = & \eta \, e^{-ip_{1}\wedge p_{2}} \, a^{\dagger}_{p_{2}} a_{p_{1}} + (2\pi)^3 \, 2 E_{p_1} \,\delta^{3}(p_{1} \, - \, p_{2}),
\label{tcom}
\end{eqnarray}
where $E^2_p = \vec{p}^2 + m^2 $, $p \wedge q \, = \, p_\mu \theta^{\mu\nu} q_\nu$ and $\eta = \pm 1$ depending on whether the particles are ``twisted bosons'' $(+1)$ or 
``twisted fermions'' $(-1)$. Because of (\ref{tcom}) the quantum fields written on GM plane, unlike ordinary quantum fields, follow an unusual statistics called twisted statistics. 
Noncommutative field theories without twisted commutation relations do not preserve the classical twisted Poincar\'e invariance at quantum level and suffer from UV/IR mixing 
\cite{pinzul}. The twisted statistics is a novel feature of fields on GM plane. It leads to interesting new effects like Pauli forbidden transitions \cite{bal,pramod} and changes 
in certain thermodynamic quantities \cite{basu, basu1}. It can be used to search for signals of noncommutativity in certain experiments involving 
U.H.E.C.Rs \cite{rahul} and C.M.B \cite{anosh}.
 
Twisted operators of (\ref{tcom}), can be used to construct noncommutative fields. For instance, a real scalar field $\phi_{\theta} $ has a normal mode expansion of the form 
\begin{eqnarray}
 \phi_{\theta} (x) & = & \int \frac{d^3 k}{(2\pi)^3 2 E_k}  \left[ a_k \, e^{-ikx} \, + \, a^\dagger_k \, e^{ikx}  \right].
\label{ncfield}
\end{eqnarray}
Using the twisted fields one can write field theories on GM plane. Twisted field theories involving real scalar field $\phi_{\theta}$ and having a $\phi^4_{\theta,\ast}$ interactions 
are discussed in \cite{bal-uvir} and are shown to be free from UV/IR mixing. Gauge field theories with nonabelian gauge groups are constructed in \cite{amilcar-pinzul, sachin-pinzul}. Construction of thermal field theories is done in \cite{amilcar-th1, amilcar-th2, trg } while \cite{asrarul} discusses the twisted bosonization in two dimensional noncommutative spacetime. A comprehensive review of twisted field theories can be found in \cite{balreview}.

The twisted creation/annihilation operators ($a^{\dagger}_p, a_p $) are related to ordinary \, \, \, \, \, \, \, \, \, creation/annihilation operators ($ c^{\dagger}_p, c_p $) 
satisfying usual statistics by the ``dressing transformation'' : 
\begin{eqnarray}
a_p & = &  c_p \, e^{-\frac{i}{2} \, p \wedge P}, \nonumber \\
a^\dagger_p & = &  c^\dagger_p \, e^{\frac{i}{2} \, p \wedge P}, 
\label{dresstransform}
\end{eqnarray}
where $ P_\mu \, = \, \int \frac{d^3 p}{(2\pi)^3 2 E_p} \, p_\mu \, c^\dagger_p c_p \, = \, \int  \frac{d^3 p}{(2\pi)^3 2 E_p} \, p_\mu a^\dagger_p a_p $ is the 
Fock space momentum operator. Using the ``dressing transformation'' of (\ref{dresstransform}), one can relate $\phi_{\theta}$ with the commutative real scalar field $\phi_{0} $ as
\begin{eqnarray}
 \phi_{\theta} (x) & = & \phi_{0} (x) \, e^{\frac{1}{2} \, \overleftarrow{\partial} \wedge P}.
\label{fielddress}
\end{eqnarray}
This is an important identity and helps us to relate noncommutative expressions with their analogous commutative ones. In what follows we will repeatedly make use of the relations 
(\ref{tcom})-(\ref{fielddress}) to simplify our computations. 

In this paper, we show that any generic (polynomial interaction terms) noncommutative field theory with only matter fields is a renormalizable theory, provided the corresponding commutative theory is also 
renormalizable. Moreover, we show that all such theories are free of UV/IR mixing. We further argue that they have identical fixed points as analogous commutative theory. 
We also obtain the $\beta$-functions for the various couplings in analogy with commutative theory.

The plan of the paper is as follows. We first start reviewing the formalism of noncommutative interaction picture and the noncommutative scattering theory. For the sake of simplicity 
we choose a specific model of the noncommutative real scalar fields having a $ \phi^4_{\theta,\ast}$ self interaction. We compute the S-matrix elements and show that for any given initial and final states, the S-matrix 
elements have only an overall noncommutative phase and hence absence of UV/IR mixing in this theory. Moreover, since the noncommutative S-matrix elements are related to the commutative ones only by an overall phase hence 
various physical observables like transition probabilities, cross section and decay rates etc remain same as those for the analogous commutative theory. Nonetheless, as discussed in \cite{bal, pramod, basu, basu1, rahul, anosh} 
various other collective mode phenomenons particularly those depending crucially on statistics of the particles do get changed and offer testable predictions for the noncommutative theory.  

After the interaction picture discussion of the scattering process, we review the noncommutative LSZ formalism (again for simplicity we will restrict only to real scalar fields) for computing S-matrix elements. 
We show that the LSZ approach also leads to the same results as the interaction picture approach and hence establish the equivalence of the two approaches. Moreover, we show that although the ``on-shell'' noncommutative Green's functions
are related to their commutative counterparts by overall noncommutative phases but that is not the case with ``off-shell'' Green's functions, which have more complicated dependence on noncommutative parameters.  

We then present our work on renormalization of this theory and show that it is renormalizable. We further compute the fixed point and $\beta$-function for 
the coupling. We show that this noncommutative theory shares the same fixed point and $\beta$-function as the analogous commutative $\phi^4_0$ theory. We also show the absence of 
UV/IR mixing in the renormalized theory. We then conclude with comments about more complicated noncommutative theories with generic polynomial interactions and involving only matter fields. 
We finally argue that our analysis although explicitly done only for a specific model holds true for all such theories.


\section{Noncommutative Interaction Picture}  

For the sake of completeness, in this section, we start reviewing the formalism of scattering theory for a generic noncommutative theory using the 
``noncommutative interaction picture''. For the sake of simplicity and definiteness, we choose a specific type of interaction hamiltonian 
$H_{\theta, \rm{Int}} = \phi^4_{\theta,\ast}$. We will compute the S-matrix $\hat{S}_\theta$ and S-matrix elements for a generic scattering problem. We also show the relation of 
these quantities with the commutative S-matrix $\hat{S}_0$ and S-matrix elements. The results discussed in this section are due to the work of \cite{bal-uvir} and the interested 
reader is referred to it for further details.


\subsection{General Formalism}

Field theories are usually done using the so called Dirac or interaction picture. Using interaction picture for calculations has many obvious advantages, making the calculations 
much easier. Hence, it is desirable, for the work done here, to have a noncommutative interaction picture. With this in mind, we briefly review the noncommutative interaction picture.
 The formalism developed here is quite similar to that of ordinary commutative field theories, for which any good book on field theory \cite{weinberg,greiner} can be consulted.

Let $\hat{H}_{\theta}$ be the full Hamiltonian for the system of interest and we assume that it can be split into two parts, the free part $\hat{H}_{\theta,F}$ and the interaction 
part $\hat{H}_{\theta, \rm{Int}} $ i.e.
\begin{eqnarray}
\hat{H}_{\theta} = \hat{H}_{\theta, F} + \hat{H}_{\theta, \rm{Int}}.
 \label{ham}
\end{eqnarray}
Let $\hat{O}^{H}_{\theta}(t)$ be a noncommutative operator in the Heisenberg Picture satisfying the Heisenberg equation of motion
\begin{eqnarray}
i\partial_{t}\hat{O}^{H}_{\theta}(t) = \left[\hat{O}^{H}_{\theta}(t) , \hat{H}_{\theta} \right].
 \label{heq}
\end{eqnarray}  
The formal solution of (\ref{heq}) is given by
\begin{eqnarray}
\hat{O}^{H}_{\theta}(t) = e^{i \hat{H}_{\theta}(t-t_{0})} \hat{O}^{H}_{\theta}(t_{0}) e^{-i \hat{H}_{\theta}(t-t_{0})}.
 \label{htev}
\end{eqnarray}
Furthermore, like in the commutative case, the state vectors $\left| \alpha, t \right> ^{H}_{\theta} $ are constant, i.e.
\begin{eqnarray}
 \left| \alpha, t \right> ^{H}_{\theta} = \left| \alpha, t_{0} \right> ^{H}_{\theta} \equiv \left| \alpha \right> ^{H}_{\theta}.
\label{hstate}
\end{eqnarray}
Now, we define the noncommutative interaction picture operator $\hat{O}^{I}_{\theta}(t)$ and state vector $\left| \alpha, t \right> ^{I}_{\theta} $ as
\begin{eqnarray}
\hat{O}^{I}_{\theta}(t)  =  e^{i \hat{H}_{\theta,F}t} e^{-i \hat{H}_{\theta}t} \hat{O}^{H}_{\theta}(t) e^{i \hat{H}_{\theta}t} e^{-i \hat{H}_{\theta,F}t}   
\label{iop}
\end{eqnarray}
and
\begin{eqnarray}
 \left| \alpha, t \right> ^{I}_{\theta} =  e^{i \hat{H}_{\theta,F}t} e^{-i \hat{H}_{\theta}t} \left| \alpha \right> ^{H}_{\theta}.
\label{istate}
\end{eqnarray}
In writing (\ref{iop}) and (\ref{istate}) we have assumed that the two pictures agree at the (arbitrarily chosen) time $t_0$.

The interaction picture operator $\hat{O}^{I}_{\theta}(t)$ defined by (\ref{iop}) satisfies the equation of motion
\begin{eqnarray}
i\partial_{t}\hat{O}^{I}_{\theta}(t) = \left[\hat{O}^{I}_{\theta}(t) , \hat{H}_{\theta,F} \right]
 \label{ieq}
\end{eqnarray} 
with formal solution written as
\begin{eqnarray}
\hat{O}^{I}_{\theta}(t) = e^{i \hat{H}_{\theta,F}(t-t_{0})} \hat{O}^{I}_{\theta}(t_{0}) e^{-i \hat{H}_{\theta,F}(t-t_{0})}.
 \label{itev}
\end{eqnarray}

Also, the state vectors $\left| \alpha, t \right> ^{I}_{\theta} $ defined by (\ref{istate}) satisfy
\begin{eqnarray}
 i \partial_t \left| \alpha, t \right> ^{I}_{\theta} =  \hat{H}^{I}_{\theta, \rm{Int}} \left| \alpha,t \right> ^{I}_{\theta}.
\label{iseveq}
\end{eqnarray}
The formal solution of (\ref{iseveq}) is given by 
\begin{eqnarray}
\left| \alpha, t \right> ^{I}_{\theta}  &  =  &  \hat{U}_{\theta}(t , t_{0})  \left| \alpha,t_{0} \right> ^{I}_{\theta}  \nonumber \\ 
& = &  e^{i \hat{H}_{\theta,F}t}  e^{-i \hat{H}_{\theta}(t - t_{0})}  e^{-i \hat{H}_{\theta,F}t_{0}}  \left| \alpha,t_{0} \right> ^{I}_{\theta}
\label{istev}
\end{eqnarray}

The operator $ \hat{U}_{\theta}(t , t_{0}) $ is the ``noncommutative time evolution operator''.  Just like its commutative counterpart it also satisfies certain properties :

\begin{enumerate}
\item  Group Property:
\begin{eqnarray}
 \hat{U}_{\theta}(t_{2} , t_{1}) \hat{U}_{\theta}(t_{1} , t_{0}) =  \hat{U}_{\theta}(t_{2} , t_{0}).
\label{tg}
\end{eqnarray}

\item Identity:
\begin{eqnarray}
 \hat{U}_{\theta}(t_{0} , t_{0})  =  \mathbb{I}.
\label{tid}
\end{eqnarray}

\item Inverse Operator:
\begin{eqnarray}
 \hat{U}^{-1}_{\theta}(t_{1} , t_{0}) =  \hat{U}_{\theta}(t_{0} , t_{1}).
\label{tin}
\end{eqnarray}

\item Unitarity:
\begin{eqnarray}
 \hat{U}^{\dagger}_{\theta}(t_{1} , t_{0}) =  \hat{U}^{-1}_{\theta}(t_{1} , t_{0}).
\label{tun}
\end{eqnarray}

\item Relation between Heisenberg and interaction pictures: If the two pictures agree at (an arbitrarily chosen) time $t = t_0$ , then we have
\begin{eqnarray}
 \hat{O}^{I}_{\theta}(t) =  \hat{U}_{\theta}(t , t_0) \hat{O}^{H}_{\theta}(t) \hat{U}^{\dagger}_{\theta}(t , t_0)
\label{tevop}
\end{eqnarray}
and
\begin{eqnarray}
  \left| \alpha, t \right> ^{I}_{\theta} =  \hat{U}_{\theta}(t , t_0)   \left| \alpha\right> ^{H}_{\theta},
\label{tevs}
\end{eqnarray}
so that $ \hat{U}_{\theta}(t , t_{0}) $ satisfies the differential equation
 \begin{eqnarray}
 i \partial_{t} \hat{U}_{\theta}(t , t_{0}) =   \hat{H}^{I}_{\theta, \rm{Int}} (t) \hat{U}_{\theta}(t , t_{0})
\label{tdfeq}
 \end{eqnarray}
with the boundary condition given by (\ref{tid}). This differential equation can be transformed into an equivalent integral equation, in  exactly the same manner 
as done in commutative field theory and we have 
\begin{eqnarray}
 \hat{U}_{\theta}(t , t_{0}) \quad = \quad \mathbb{I} \quad + \quad  (-i) \int_{t_{0}}^{t} dt' \hat{H}^{I}_{\theta, \rm{Int}} (t') \hat{U}_{\theta}(t' , t_{0}).
\label{tinteq}
\end{eqnarray}
The formal solution of  (\ref{tinteq}) can be written in terms of ``time ordered exponential function'' as
\begin{eqnarray}
  \hat{U}_{\theta}(t , t_{0}) = \mathcal{T} \exp\left[-i\int^{t}_{t_{0}} dt' \hat{H}^{I}_{\theta, \rm{Int}} (t')\right]
\label{tev}
\end{eqnarray}  
where the time ordering operator $\mathcal{T}$ is defined in the same way as in standard commutative case. 
\end{enumerate}


\subsection{Computation of S-matrix}

In the previous section we have developed the noncommutative interaction picture. In this section we use it to compute S-matrix elements for a typical scattering process. 
We use a particular model of real scalar fields having quartic self-interactions. The commutative interaction Hamiltonian density $\mathcal{\hat{H}}_{0, \rm{Int}}(x)$ that we 
consider is given by
\begin{eqnarray}
\mathcal{\hat{H}}_{0, \rm{Int}}(x) & = & \frac{\lambda}{4!} \, \phi_0 (x) \cdot \phi_0 (x) \cdot \phi_0 (x) \cdot \phi_0 (x) \, = \,  \frac{\lambda}{4!} \, \phi^4_0 (x)
\label{chint}
\end{eqnarray}
and the analogous noncommutative interaction hamiltonian density $\mathcal{\hat{H}}_{\theta, \rm{Int}}(x)$ is 
\begin{eqnarray}
 \mathcal{\hat{H}}_{\theta, \rm{Int}} (x) & = & \frac{\lambda}{4!} \, \phi_{\theta} (x) \ast \phi_{\theta}(x) \ast \phi_{\theta}(x) \ast \phi_{\theta} (x) 
\, = \, \frac{\lambda}{4!} \, \phi^4_{\theta,\ast} (x) \, = \, \frac{\lambda}{4!} \, \phi^4_0 (x) \, e^{\frac{1}{2} \overleftarrow{\partial} \wedge P},
\label{nchint}
\end{eqnarray}
where in writing the last equality we have used the dressing transformation (\ref{fielddress}) and the expression for the star product (\ref{starproduct}). 

Our aim is to compute the noncommutative S-matrix elements for a typical scattering process. We do that by first finding a relation between noncommutative S-matrix elements and 
their commutative counterparts by making use of the dressing transformations (\ref{dresstransform}) and (\ref{fielddress}). We briefly review the standard treatment in commutative 
case before discussing the noncommutative case and establishing its relation with commutative case.


\subsubsection{Commutative Case}

Let us restrict ourselves to two particles scattering processes $ p_{1}, p_{2} \rightarrow  p'_{1}, p'_{2} $. The case of two-to-many and many-to-many will be 
taken up later. For a typical two-to-two particle scattering, the S-matrix element is given by
\begin{eqnarray}
 S_{0}[p_{2}, p_{1} \rightarrow  p'_{1}, p'_{2}] & \equiv &  S_{0}[p'_{2}, p'_{1} ; p_{2}, p_{1}]
 \; = \;  \leftidx{_{\rm{out},0}}{\left \langle p'_{2}, p'_{1} | p_{2}, p_{1} \right \rangle }{_{0,\rm{in}}}
\label{csm}
\end{eqnarray}
where $ |p'_{1}, p'_{2} \rangle_{0,\rm{out}} $ is the two particle out-state measured in the far future and $ |p_{2}, p_{1} \rangle_{0,\rm{in}} $ is the two particle in-state 
prepared in the far past. The in- and out-states can be related with each other using S-matrix $\hat{S}_{0}$. Therefore we have
\begin{eqnarray}
S_{0}[p'_{2}, p'_{1} ; p_{2}, p_{1}] & = &  \leftidx{_{\rm{out},0}}{\left \langle  p'_{2}, p'_{1}| \hat{S}_{0} | p_{2}, p_{1}  \right \rangle}{_{\rm{out},0}}
\; = \; \leftidx{_{\rm{in},0}}{\left \langle  p'_{2}, p'_{1}| \hat{S}_{0} | p_{2}, p_{1}  \right \rangle}{_{\rm{in},0}}.
\label{csmat}
\end{eqnarray}
In the interaction picture $\hat{S}_{0}$ can be written as 
\begin{eqnarray}
 \hat{S}_{0} & = & \lim_{t_{1} \to +\infty \atop t_{2} \to -\infty} U_{0}(t_{1},t_{2}) \nonumber \\
& = & \mathcal{T} \exp\left[-i\int^{\infty}_{-\infty} d^{4}z \frac{\lambda}{4!} \phi^{4}_{0}(z) \right].
\label{csoperator}
\end{eqnarray}
In the last line we have used the form (\ref{chint}) for the interaction Hamiltonian density.

Also, in the interaction picture, the two particle states are defined as
\begin{eqnarray}
  |p, q \rangle_{0} & = & c^{\dagger}_{q} c^{\dagger}_{p}  |0 \rangle
\label{c2state}
\end{eqnarray}
where $ c^{\dagger}_{p} $ is the interaction picture creation operator for the commutative theory with the usual commutation relations.

Using (\ref{csoperator}) and (\ref{c2state}) we obtain 
\begin{eqnarray}
 S_{0}[p'_{2}, p'_{1} ; p_{2}, p_{1}] & = &  \lim_{t_1 \to +\infty \atop t_2 \to -\infty} \left \langle 0 \left | c_{p'_{1}} c_{p'_{2}} \mathcal{T} \exp\left[-i\int^{t_1}_{t_2} 
d^{4}z \frac{\lambda}{4!} \phi^{4}_{0}(z) \right] c^{\dagger}_{p_{1}} c^{\dagger}_{p_{2}} \right |0 \right \rangle.
\label{csmatrix}
\end{eqnarray}
Now, to calculate any specific process, $\hat{S}_{0}$ is expanded in power series of coupling constant $\lambda$ (provided $\lambda$ is small enough to allow
 perturbative expansion) up to some desired order of coupling constant. It is evaluated using standard techniques, e.g. Wick's theorem and Feynman diagrams. 

The two-to-many ($ 2 \rightarrow N $) or many-to-many particle ($ M \rightarrow N $)  scattering cases can be similarly discussed. For instance, 
for ($ M \rightarrow N $) scattering we have
\begin{eqnarray}
 S_{0}[ p'_{N}, ... p'_{1} ; p_{M}, ... p_{1} ] & = &  \leftidx{_{\rm{out},0}}{\left \langle p'_{N}, ... p'_{1} | p_{M} ... p_{1} \right \rangle }{_{0,\rm{in}}}
\label{cmsm}
\end{eqnarray}
where $ |p'_{1}, ... p'_{N} \rangle_{0,\rm{out}} $ is the N-particle out-state and $ | p_{M} ... p_{1} \rangle_{0,\rm{in}} $ is the M-particle in-state.
 As before, the in- and out-states can be related with each other using S-matrix $\hat{S}_{0}$. Therefore we have
\begin{eqnarray}
 S_{\theta}[ p'_{N}, ... p'_{1} ; p_{M}, ... p_{1} ]  =  \leftidx{_{\rm{out},0}}{\left \langle p'_{N}, ... p'_{1} \left| \hat{S}_{0} \right| p_{M} ... p_{1}
 \right \rangle}{_{\rm{out},0}}
 =  \leftidx{_{\rm{in},0}}{\left \langle p'_{N}, ... p'_{1} \left| \hat{S}_{0} \right| p_{M} ... p_{1} \right \rangle}{_{\rm{in},0}}.
\label{cmsmat}
\end{eqnarray}

In the interaction picture, $\hat{S}_{0}$ is given by $(\ref{csoperator})$ and the multiple-particle states can be written as 
\begin{eqnarray}
  |p_{M} ... p_{1} \rangle_{0} & = & c^{\dagger}_{p_{1}} ... c^{\dagger}_{p_{M}} |0 \rangle.
\label{cmstate}
\end{eqnarray}
Using (\ref{csoperator}) and (\ref{cmstate}) we obtain
\begin{equation}
  S_{0}[ p'_{N}, ... p'_{1} ; p_{M}, ... p_{1} ]   =  \lim_{t_{1} \to +\infty \atop t_{2} \to -\infty} \left \langle 0 \left | c_{p'_{1}} ... c_{p'_{N}} 
\mathcal{T} \exp\left[-i\int^{t_1}_{t_2} d^{4}z \frac{\lambda}{4!} \phi^{4}_{0}(z) \right] c^{\dagger}_{p_{1}} ... c^{\dagger}_{p_{M}} \right |0 \right \rangle.
\label{cmsmatrix}
\end{equation}
Again, any specific process can be calculated using perturbative expansion in $\lambda$ (if possible) and invoking standard tools like Wick's theorem and Feynman diagrams.


\subsubsection{Noncommutative Case}

Our treatment of the noncommutative case follows closely the formalism of commutative case. Therefore, as in the commutative case, for a two-to-two particle scattering processes 
the S-matrix elements are given by
\begin{eqnarray}
 S_{\theta}[p_{2}, p_{1} \rightarrow  p'_{1}, p'_{2}] & \equiv &  S_{\theta}[p'_{2}, p'_{1} ; p_{2}, p_{1}]
\; = \;  \leftidx{_{\rm{out},\theta}}{\left \langle  p'_{2}, p'_{1} | p_{2}, p_{1} \right \rangle }{_{ \theta, \rm{in}}}
\label{ncsm}
\end{eqnarray}
where $ |p'_{1}, p'_{2} \rangle_{\theta,\rm{out}} $ is the noncommutative two particle out-state which is measured in the far future and $ |p_{2}, p_{1}  \rangle_{\theta,\rm{in}} $ 
is the noncommutative two particle in-state prepared in the far past. Now, because of the twisted statistics (\ref{tcom}) there is an ambiguity in defining the action of the 
twisted creation and annihilation operators on the Fock space of states. Following \cite{bal-pinzul} we choose to define $a^\dagger_k$ to be an operator which adds a particle to 
the right of the particle list,
\begin{eqnarray}
 a^\dagger_k | p_1,p_2 \dots p_n \rangle_\theta & = & | p_1,p_2 \dots p_n, k \rangle_\theta.
\label{taanni}
\end{eqnarray}
Hence the two particle in-state can be written as 
\begin{eqnarray}
 |p_{2}, p_{1}  \rangle_{\theta,\rm{in}} & = & a^\dagger_{p_1}\, a^\dagger_{p_2} \, | 0 \rangle.
\label{int2state}
\end{eqnarray}
Since the noncommutative vacuum is the same as that of the commutative theory, no extra label is needed for $| 0 \rangle$.

Just like in the commutative case, the noncommutative in- and out-states can be related with each other using S-matrix $\hat{S}_{\theta}$. Therefore we have
\begin{eqnarray}
S_{\theta}[p'_{2}, p'_{1} ; p_{2}, p_{1}] & = & \leftidx{_{\rm{out},\theta}}{\left \langle p'_{2}, p'_{1} \left|\hat{S}_{\theta} \right| p_{2}, p_{1} \right \rangle}{_{\rm{out},\theta}}
\; = \;  \leftidx{_{\rm{in},\theta}}{\left \langle  p'_{2}, p'_{1} \left | \hat{S}_{\theta} \right| p_{2}, p_{1}  \right \rangle}{_{\rm{in},\theta}}.
\label{rncsmat}
\end{eqnarray}

The noncommutative S-matrix $\hat{S}_{\theta}$ in the interaction picture can be written as 
\begin{eqnarray}
 \hat{S}_{\theta} & = & \lim_{t_{1} \to +\infty \atop t_{2} \to -\infty} U_{\theta}(t_{1},t_{2})
\label{ncsop}
\end{eqnarray} 
where $U_{\theta}(t_{1},t_{2}) $ is given by (\ref{tev}). For the interaction Hamiltonian density given in (\ref{nchint}) we obtain
\begin{eqnarray}
\hat{S}_{\theta} & = & \mathcal{T}  \exp \left[ -i\int^{\infty}_{-\infty} d^{4}z \frac{\lambda}{4!} \phi^{4}_{0}(z) e^{\frac{1}{2} \overleftarrow{\partial_{z}} \wedge P }\right]. 
\label{ncsint}
\end{eqnarray}
One can formally expand the exponential and write $\hat{S}_{\theta}$ as a time-ordered power series like
\begin{eqnarray}
\hat{S}_{\theta} & = &   \mathbb{I} \, + \, -i\int^{\infty}_{-\infty} d^{4}z   \frac{\lambda}{4!} \phi^{4}_{0}(z) \, e^{\frac{1}{2} \overleftarrow{\partial_{z}} \wedge P } \nonumber \\
& + &  \mathcal{T}\, (-i )^2 \int^{\infty}_{-\infty} d^{4}z \int^{\infty}_{-\infty} d^{4}z'\,  \frac{\lambda}{4!} \phi^{4}_{0}(z) \, e^{\frac{1}{2} \overleftarrow{\partial_{z}} \wedge P } 
\,  \frac{\lambda}{4!} \phi^{4}_{0}(z') \, e^{\frac{1}{2} \overleftarrow{\partial_{z'}} \wedge P } \, + \, \cdots
\label{rspower}
\end{eqnarray} 
As done in \cite{bal-uvir}, each term in the power series in (\ref{rspower}) can be further simplified by expanding the exponential $ e^{\frac{1}{2} \overleftarrow{\partial} \wedge P } $, 
integrating by parts and discarding the surface terms. For instance, the second term in (\ref{rspower}) becomes
\begin{eqnarray}
 -i\int^{\infty}_{-\infty} d^{4}z    \frac{\lambda}{4!} \phi^{4}_{0}(z) \, e^{\frac{1}{2} \overleftarrow{\partial_{z}} \wedge P } 
& = &  -i\int^{\infty}_{-\infty} d^{4}z \left [   \frac{\lambda}{4!} \phi^{4}_{0}(z) 
\, + \, \partial_\mu \left( \frac{\lambda}{4!} \phi^{4}_{0}(z) \right)\, \theta^{\mu \nu}\, P_\nu \, + \, \dots \right] \nonumber \\
& = & -i\int^{\infty}_{-\infty} d^{4}z \frac{\lambda}{4!} \phi^{4}_{0}(z) .
\label{rsecpower}
\end{eqnarray}
One can similarly show that all the higher order terms in the power series of (\ref{rspower}) are also free 
of any $\theta$ dependence. We refer to \cite{bal-uvir} for more details.

We then have
\begin{eqnarray}
 \hat{S}_{\theta} & = & \mathcal{T} \exp\left[-i\int^{\infty}_{-\infty} d^{4}z \frac{\lambda}{4!} \phi^{4}_{0}(z) \right] \; = \; \hat{S}_{0}.
\label{rncsoperator}
\end{eqnarray}
Using (\ref{rncsoperator}) and (\ref{int2state}), the $S$-matrix elements can be written as 
\begin{eqnarray}
 S_{\theta}[p'_{2}, p'_{1} ; p_{2}, p_{1}] & = & \lim_{t_{1} \to +\infty \atop t_{2} \to -\infty}\left \langle 0 \left | a_{p'_{1}} a_{p'_{2}} \mathcal{T} \exp\left[-i\int^{t_1}_{t_2} 
d^{4}z \frac{\lambda}{4!}  \phi^{4}_{0}(z) \right] a^{\dagger}_{p_{1}} a^{\dagger}_{p_{2}} \right |0 \right \rangle.
\label{ncsmatr}
\end{eqnarray}
But the noncommutative creation/annihilation operators are related with those of commutative theory by dressing transformation (\ref{dresstransform}), so that
\begin{eqnarray}
 S_{\theta}[p'_{2}, p'_{1} ; p_{2}, p_{1}] & = & \lim_{t_{1} \to +\infty \atop t_{2} \to -\infty} \left \langle 0 \left | c_{p'_{1}} e^{\frac{-i}{2} p'_{1} \wedge P} c_{p'_{2}} 
e^{\frac{-i}{2} p'_{2} \wedge P}  \mathcal{T} \exp\left[-i\int^{t_1}_{t_2} d^{4}z \frac{\lambda}{4!} \phi^{4}_{0}(z) \right] \right. \right. \nonumber \\
& & \left. \left. c^{\dagger}_{p_{1}} e^{\frac{i}{2} p_{1} \wedge P} c^{\dagger}_{p_{2}} e^{\frac{i}{2} p_{2} \wedge P}\right |0 \right \rangle \nonumber \\
& = &  e^{\frac{-i}{2} p'_{2} \wedge p'_{1}} \, e^{\frac{i}{2} p_{1} \wedge p_{2}} \lim_{t_{1} \to +\infty \atop t_{2} \to -\infty} \left \langle 0 \left | c_{p'_{1}} c_{p'_{2}} 
\mathcal{T} \exp\left[-i\int^{t_1}_{t_2} d^{4}z \frac{\lambda}{4!} \phi^{4}_{0}(z) \right]   c^{\dagger}_{p_{1}} c^{\dagger}_{p_{2}} \right |0 \right \rangle \nonumber \\
& = &  e^{\frac{-i}{2} p'_{2} \wedge p'_{1}} \, e^{\frac{i}{2} p_{1} \wedge p_{2}} \; S_{0}[p'_{2}, p'_{1} ; p_{2}, p_{1}].
\label{ncsmatrix}
\end{eqnarray}
The expression (\ref{ncsmatrix}) relates the noncommutative S-matrix element for a two-to-two particle scattering process with its commutative counterpart. We remark that this 
correspondence is a nonperturbative one in $\theta$ and it is true to all orders in perturbation of the coupling constant. Also, the only noncommutative dependence of 
$S_{\theta}[p'_{2}, p'_{1} ; p_{2}, p_{1}]$ is by an overall phase. Therefore there are no non-planar diagrams and hence the model is essentially free from any UV/IR mixing.

An analogous relation between noncommutative and commutative S-matrix for two-to-many ($ 2 \rightarrow N $) and many-to-many ($ M \rightarrow N $) particle scattering processes 
can be established in a similar way. For instance, for  ($ M \rightarrow N $) scattering we have
\begin{eqnarray}
 S_{\theta}[ p'_{N}, ... p'_{1} ; p_{M}, ... p_{1} ] & = &  \leftidx{_{\rm{out},\theta}}{\left \langle  p'_{N}, ... p'_{1} | p_{M} ... p_{1} \right \rangle }{_{\theta,\rm{in}}},
\label{ncmsm} 
\end{eqnarray}
where $ |p'_{1}, ... p'_{N} \rangle_{\theta,\rm{out}} $ is the noncommutative N-particle out-state and $ | p_{M} ... p_{1} \rangle_{\theta,\rm{in}} $ is the noncommutative 
N-particle in-state. As before, the in- and out-states can be related with each other using S-matrix $\hat{S}_{\theta}$. Therefore we have
\begin{eqnarray}
 S_{\theta}[ p'_{N}, ... p'_{1} ; p_{M}, ... p_{1} ]  =  \leftidx{_{\rm{out},\theta}}{\left \langle p'_{N}, ... p'_{1} \left| \hat{S}_{0} \right| p_{M} ... p_{1}
 \right \rangle}{_{\rm{out},\theta}}
 =  \leftidx{_{\rm{in},\theta}}{\left \langle p'_{N}, ... p'_{1} \left| \hat{S}_{0} \right| p_{M} ... p_{1} \right \rangle}{_{\rm{in},\theta}}
\label{ncmsmat}
\end{eqnarray}

As before, the interaction picture noncommutative $S$-matrix $\hat{S}_{\theta}$ is given by $(\ref{rncsoperator})$. Moreover, 
just like the two-particle states, the interaction picture noncommutative multiple-particle states can be written as 
\begin{eqnarray}
  | p_{M} ... p_{1} \rangle_{\theta} & = & a^{\dagger}_{p_{1}} ... a^{\dagger}_{p_{M}}  |0 \rangle.
\label{ncmstate}
\end{eqnarray}
Using (\ref{rncsoperator}) and (\ref{ncmstate}) we obtain 
\begin{eqnarray}
S_{\theta}[ p'_{N}, ... p'_{1} ; p_{M}, ... p_{1} ] & = & \lim_{t_{1} \to +\infty \atop t_{2} \to -\infty} \left \langle 0 \left | a_{p'_{1}} ... a_{p'_{N}}
 \mathcal{T} \exp\left[-i\int^{t_1}_{t_2} d^{4}z \frac{\lambda}{4!} \phi^{4}_{0}(z) \right] a^{\dagger}_{p_{1}} ... a^{\dagger}_{p_{M}} \right |0 \right \rangle.
\label{ncmsmatr}
\end{eqnarray}
Using the dressing transformation (\ref{dresstransform}) in (\ref{ncmsmatr}) we obtain
\begin{eqnarray}
S_{\theta}[p'_{N}, ... p'_{1} ; p_{M}, ... p_{1}]  & = & \lim_{t_{1} \to +\infty \atop t_{2} \to -\infty} \left \langle 0 \left | c_{p'_{1}} e^{\frac{-i}{2} p'_{1} \wedge P} ... 
c_{p'_{N}} e^{\frac{-i}{2} p'_{N} \wedge P}  \mathcal{T} \exp\left[-i\int^{t_1}_{t_2} d^{4}z \frac{\lambda}{4!} \phi^{4}_{0}(z) \right]  \right. \right. \nonumber \\
& & \left. \left. c^{\dagger}_{p_{1}} e^{\frac{i}{2} p_{1} \wedge P} ...  c^{\dagger}_{p_{M}} e^{\frac{i}{2} p_{M} \wedge P} \right |0 \right \rangle \nonumber \\
& = &  e^{\frac{i}{2} \left( \sum^{M}_{i,j=1,j>i}p_{i} \wedge p_{j}  -   \sum^{N}_{i,j= N, j< i} p'_{i} \wedge p'_{j}\right)}  S_{0}[p'_{N}, ... p'_{1} ; p_{M}, ... p_{1}].
\label{ncmsmatrix}
\end{eqnarray}
This is the generic result relating the noncommutative many-to-many particle S-matrix with its commutative analogue. Again, it should be noted that the proof is completely 
nonperturbative in $\theta$ and hence valid to all orders in the coupling constant. Also, as argued before, the phenomena of UV/IR mixing is completely absent. Moreover, 
since the noncommutative S-matrix elements are related to the analogous commutative ones only by an overall phase, so physical observables like transition probabilities, cross section and decay rates etc 
remain unchanged. In spite of this, various other collective mode phenomenons, particularly those depending crucially on statistics of the particles do get changed and offer testable predictions for the noncommutative theory 
\cite{bal, pramod, basu, basu1, rahul, anosh}.


\section{Noncommutative LSZ Formalism}

In this section we review the noncommutative LSZ formalism and calculate the noncommutative S-matrix elements via the reduction formula.
 The noncommutative S-matrix computed via LSZ will be shown to be completely equivalent to that computed in the previous section using interaction picture.  
This establishes the equivalence  of the two approaches. Also, this second method brings out the difference between scattering amplitudes and off-shell Green's functions.

We consider as an example the time ordered product of four real scalar fields with $\phi^4$ type self-interactions representing a process of two particles going 
into two other particles. This is described by the correlation function 
\begin{eqnarray}
G_{2+2}(x_1', x_2';~x_1,x_2)=\langle\Omega|\mathcal{T}\left(\phi(x_1')\phi(x_2')\phi(x_1)\phi(x_2)\right)|\Omega\rangle
\label{gengreen}
\end{eqnarray}
where $|\Omega\rangle$ is the vacuum of the full interacting theory.

The Green's function $G_{2+2}^0$ in the commutative case is given by the time ordered product of four commutative fields $\phi_0$. The corresponding Green's function 
$G^{\theta}_{2+2}$
in the noncommutative case is obtained by replacing the commutative fields $\phi_0$ by the noncommutative ones  $\phi_\theta$ in the time ordered product in (\ref{gengreen}). 
The case of many particle scattering will be taken up later.

As done in previous section, we start first by briefly reviewing the derivation of commutative LSZ reduction formula before going on to the noncommutative case. 
The derivation presented in this section is originally due to \cite{amilcar} which can be consulted for further details.


\subsection{ Commutative Case }

In this section we use the following notations: 
\begin{eqnarray}
& & \hat{p}~ \textrm{is an on-shell momentum}= (E_{\vec{p}}=\sqrt{\vec{p}^{~2}+m^2},~ \vec{p}), \nonumber \\
& & p~ \textrm{is a generic 4-momentum, with}~ p^0>0.
\end{eqnarray}

Let us consider the time ordered product of four commutative fields $\phi_0 (x)$ given by
\begin{eqnarray}
G^0_{2+2}(x_1', x_2';~x_1,x_2) & = & \langle\Omega|\mathcal{T}\left(\phi_0(x_1')\phi_0(x_2')\phi_0(x_1)\phi_0(x_2)\right)|\Omega\rangle.
\label{com2green}
\end{eqnarray}
As mentioned before, $G^0_{2+2}(x_1', x_2';~x_1,x_2) $ can be related to a process of two particles scattering/decaying into two other particles.

We Fourier transform $G_{2+2}^0(x_1', x_2';~x_1, x_2)$ only in $x_1'$. Without loss of generality, we can assume that $x_1'$ is associated with an outgoing particle. 
We can split the $x_1'^0$-integral into three time intervals as
 \begin{eqnarray}
\left( \int_{-\infty}^{T_-}dx_1'^0 + \int_{T_-}^{T_+}dx_1'^0 + \int_{T_+}^{\infty}dx_1'^0 \right)d^3x_1'~e^{ip_1'^0x_1'^0 - i \vec{p}_1'\cdot\vec{x}_1'} \, G_{2+2}^0(x_1', x_2';~x_1, x_2)
\end{eqnarray}
Here $T_+>> \textrm{max}(x_2'^0, x_1^0, x_2^0)$ and $T_-<<\textrm{min}(x_2'^0, x_1^0, x_2^0)$. Since $T_+\geq x_1'^0 \geq T_-$ is a finite interval, the corresponding integral 
gives no pole. A pole comes from a single particle insertion in the integral over $x_1'^0\geq T_+$ in $G_{2+2}^0$. In the integration between the limits 
$T_+$ and $+\infty$, $\phi(x_1')$ stands to the extreme left inside the time-ordering so that
 \begin{eqnarray}
G_{2+2}^0(x_1', x_2';~x_1, x_2)= \int\frac{d^3q_1}{(2\pi)^3}\frac{1}{2E_{\vec{q}_1}} \langle\Omega|\phi_0(x_1')|q_1\rangle\langle q_1|
T\left(\phi_0(x_2')\phi_0(x_1)\phi_0(x_2)\right)|\Omega\rangle \, + \, \textrm{OT}
\end{eqnarray}
where $\textrm{OT}$ stands for the other terms. The matrix element of the field $\phi_0(x_1')$ can be written as
\begin{eqnarray}
 \langle\Omega|e^{iP\cdot x_1'}\phi_0(0)e^{-iP\cdot x_1'}|E_{\vec{q}_1}, \vec{q_1}\rangle & = & \langle\Omega|\phi_0(0)|E_{\vec{q_1}}, 
\vec{q_1}\rangle e^{-iq_1\cdot x_1'}|_{q_1^0=E_{\vec{q}_1}}  \nonumber \\ 
& = & \langle\Omega|\phi_0(0)|q^0_1,\vec{q}_1=0\rangle e^{-iq_1\cdot x_1'}|_{q_1^0=E_{\vec{q}_1}}
\end{eqnarray} 
where $E^2_{\vec{q_1}} = \vec{q_1}^2+m^2 $.  We have used the Lorentz invariance of the vacuum $|\Omega\rangle$ and $\phi_0(0)$ in above. We then have
\begin{eqnarray}
 \langle\Omega|\phi_0(x_1')|E_{\vec{q_1}}, \vec{q}_1\rangle = \sqrt{Z} e^{-i\left(E_{\vec{q_1}}x_1'^0-\vec{q}_1\cdot\vec{x}_1'\right)}
\end{eqnarray}
where the field-strength renormalization factor $\sqrt{Z}$ is defined by
\begin{eqnarray}
\sqrt{Z} = \langle\Omega|\phi_0(0)|q^0_1,\vec{q}_1=0\rangle
\end{eqnarray}
and $q_1^0 > 0$. Hence the integral between $T_+$ and $+\infty$ becomes
\begin{eqnarray}
\sqrt{Z}\frac{1}{2E_{\vec{p'}_1}}\int_{T_+}^{\infty}dx_1'^0 ~e^{i\left(p_1'^0 - E_{\vec{p}_1'} +i\epsilon\right)x_1'^0}~\langle p_1'|
T\left(\phi_{2'}\phi_1\phi_2\right)|\Omega\rangle \, + \, \textrm{OT}
\end{eqnarray}
where $\epsilon > 0$ is a cut-off and $\phi_i = \phi_0(x_i)$. After the $x_1'^0$ integral we obtain
\begin{eqnarray}
\widetilde{G}_0^{(1)}(p_1', x_2', x_1, x_2)= \sqrt{Z}\frac{i}{2E_{\vec{p'}_1}} \frac{e^{i\left(p_1'^0-E_{\vec{p}_1'}+i\epsilon\right)T_+}}{\left(p_1'^0-E_{\vec{p}_1'}+i\epsilon\right)}\langle
 p_1'|T\left(\phi_{2'}\phi_1\phi_2\right)|\Omega\rangle  \, + \, \textrm{OT}.
\end{eqnarray}
As $p_1'^0 \rightarrow E_{\vec{p}_1'}$, it becomes
\begin{eqnarray}
\label{pole}
\widetilde{G}_0^{(1)}(p_1', x_2', x_1, x_2) = \sqrt{Z}  \frac{i}{p_1'^2-m^2-i\epsilon}~\langle p_1'|T\left(\phi_{2'}\phi_1\phi_2\right)|\Omega\rangle \, + \, \textrm{OT}.
\end{eqnarray}
Now in the case of integration over $(-\infty, T_-)$, $\phi_0(x_1')$ stands to the extreme right in the time ordered product, so the one-particle state contribution comes from
\begin{eqnarray}
 \langle q_1|\phi_0(x_1')|\Omega\rangle = \sqrt{Z}e^{i\left(E_{\vec{q}_1}x_1'^0-\vec{q}_1\cdot\vec{x}_1'\right)}.
\end{eqnarray}
The energy denominator is thus $\frac{1}{p_1'^0+E_{\vec{p}_1'}-i\epsilon}$ and has no pole for $p_1'^0>0$. The only pole comes from the single particle insertion in the integral 
over $x_1'^0\geq T_+$. It is given by (\ref{pole}).

Similarly, for the two-particle scattering $p_1, p_2\rightarrow p_1', p_2'$, the poles appear in both $p_1'^0$ and $p_2'^0$ when both $x_1'^0$ and $x_2'^0$ integrations are large, 
that is
\begin{eqnarray}
  x_1'^0,~ x_2'^0~ >>~ T_1~>>~x_1^0,~ x_2^0.
\end{eqnarray}
So for these poles, we obtain
\begin{eqnarray}
 \widetilde{G}_0^{(2)}(p_1', p_2', x_1, x_2) & = & \int_{T_+}^{\infty} dx_1'^0 dx_2'^0 d^3x_1' d^3x_2'~ e^{ip_1'\cdot x_1' + ip_2'\cdot x_2'}\frac{1}{2!}\left(\frac{1}{(2\pi)^3}\right)^2
\frac{d^3q_1 d^3q_2}{(2E_{\vec{q}_1})(2E_{\vec{q}_2})} \nonumber \\ 
& & \langle\Omega|\phi_0(x_1')\phi_0(x_2')|\vec{q}_1\vec{q}_2\rangle\langle\vec{q}_1\vec{q}_2|T\left(\phi_1\phi_2\right)|\Omega\rangle \, + \, \textrm{OT}.
\end{eqnarray}
Here $T_+$ is supposed to be very large. We take $\phi_0(x_1')$, $\phi_0(x_2')$ to be out fields. As we set $|\vec{q}_2\vec{q}_1\rangle$ to $|\vec{q}_2\vec{q}_1\rangle_{\textrm{out}}$
 for large $T_+$, only $\langle\Omega|\phi^{\textrm{out}+}_0(x_1')\phi^{\textrm{out}+}_0(x_2')|\vec{q}_2\vec{q}_1\rangle_{\textrm{out}}$, where $\phi_0^{\textrm{out}+}$ is 
the positive frequency part of the out-field, contributes. Thus we do not need any time-ordering between these out-fields. So we have
\begin{eqnarray}
\widetilde{G}_0^{(2)}(p_1',p_2',x_1,x_2) & = & \int_{T_+}^{\infty}d^4x_1'd^4x_2' ~e^{ip_1'\cdot x_1'+ip_2'\cdot x_2'}\frac{1}{2!}\left(\frac{1}{(2\pi)^3}\right)^2
\left(\frac{d^3q_1}{2E_{\vec{q}_1}}\right)\left(\frac{d^3q_2}{2E_{\vec{q}_2}}\right) \nonumber \\ 
& & \langle\Omega|\phi_0^{\textrm{out}}(x_1')\phi_0^{\textrm{out}}(x_2')|\vec{q}_2\vec{q}_1\rangle_{\textrm{out}}~_{\textrm{out}}\langle\vec{q}_2\vec{q}_1|T
\left(\phi_1\phi_2\right)|\Omega\rangle.
\end{eqnarray}
Now, 
\begin{eqnarray}
\langle\Omega|\phi_0^{\textrm{out}}(x_1')\phi_0^{\textrm{out}}(x_2')|\vec{q}_2\vec{q}_1\rangle_{\textrm{out}}=\langle\Omega|\phi_0^{\textrm{out}}(x_1')|\vec{q}_1
\rangle\langle\Omega|\phi_0^{\textrm{out}}(x_2')|\vec{q}_2\rangle + \vec{q}_2\leftrightarrow\vec{q}_1.
\end{eqnarray}
Thus we can generalize (\ref{pole}) to
\begin{eqnarray}
\widetilde{G}_0^{(2)}(p_1',p_2', x_1,x_2) & = & \left[\sqrt{Z} \left(\frac{i}{p^{'2}_1-m^2-i\epsilon}\right)\right]\left[\sqrt{Z} \left(\frac{i}{p^{'2}_2-m^2-i\epsilon}\right)\right] 
\nonumber \\
& &  _{\textrm{out}}\langle p_1' p_2'|T\left(\phi_1\phi_2\right)|\Omega\rangle \, + \, \textrm{OT}.
\end{eqnarray}
Similar calculations for incoming poles, with $x_1^0, x_2^0 << T_- << x_1'^0, x_2'^0$, leads to
\begin{eqnarray}
\label{commutativeLSZ}
\widetilde{G}_0^{(4)}(p_1', p_2', p_1, p_2) & = & \prod_{i=1}^2\prod_{j=1}^2 \left[\sqrt{Z}  \left(\frac{1}{p_i^{'2}-m^2-i\epsilon}\right)\right]\left[\sqrt{Z} 
\left(\frac{1}{p_j^2-m^2-i\epsilon}\right)\right] \nonumber \\
& &  _{\textrm{out}}\langle p_1'~ p_2'~|~ p_1~p_2\rangle_{\textrm{in}}.
\end{eqnarray}


\subsection{Noncommutative Case}

Our treatment of the noncommutative case is quite similar to that of the commutative case just discussed. Our aim is to arrive at the noncommutative 
version of (\ref{commutativeLSZ}). However, instead of considering a 2-particle scattering process first and then generalizing, as done in the commutative case, we directly 
start with the generic process where $M$ particles go into $N$ particles. 

Before discussing the noncommutative LSZ formalism we list down a few relations:

\begin{enumerate}
\item \textbf{The completeness relations :} These remain same for the twisted in- and out-states like in the commutative case. Recall that the noncommutative phases arising because 
of the
twisted statistics (\ref{tcom}) followed by $a_p$ and $ a^\dagger_p$,  cancel each other. Therefore
 \begin{eqnarray}
\label{5.25}
 a^{\dag\textrm{in, out}}_{p_N}\cdots a^{\dag\textrm{in, out}}_{p_1}|\Omega\rangle\langle\Omega| a^{\textrm{in, out}}_{p_1}\cdots a^{\textrm{in, out}}_{p_N}  
 =   c^{\dag\textrm{in, out}}_{p_N}\cdots c^{\dag\textrm{in, out}}_{p_1}| \Omega\rangle\langle\Omega|c^{\textrm{in, out}}_{p_1}\cdots c^{\textrm{in, out}}_{p_N}.
\end{eqnarray} 
Using (\ref{5.25}) one can also check the resolution of identity (given below) as well as the completeness for the twisted in- and out-states.

\item \textbf{ Resolution of identity:} 
\begin{eqnarray}
I' = \sum_N\frac{1}{N!}\left(\int \prod_{i=1}^N \frac{d^3p_i }{(2\pi)^3}\frac{1}{2E_{\vec{p_i}}}\right) a^{\dag\textrm{in, out}}_{p_N}\cdots a^{\dag\textrm{in, out}}_{p_1}
|\Omega\rangle\langle\Omega|a^{\textrm{in, out}}_{p_1}\cdots a^{\textrm{in, out}}_{p_N}.
\label{reid}
\end{eqnarray} 
This turns out to be independent of $\theta_{\mu\nu}$ due to (\ref{5.25}). Hence we have
\begin{eqnarray}
\label{id}
I' = \sum_N\frac{1}{N!}\left(\int \prod_{i=1}^N \frac{d^3p_i}{(2\pi)^3}\frac{1}{2E_{\vec{p_i}}}\right) c^{\dag\textrm{in, out}}_{p_N}
\cdots c^{\dag\textrm{in, out}}_{p_1}|\Omega\rangle\langle\Omega|c^{\textrm{in, out}}_{p_1}\cdots c^{\textrm{in, out}}_{p_N}.
\end{eqnarray}
\end{enumerate}

We are interested in the scattering process of $M$ particles going to $N$ particles. We then consider the twisted $N+M$-point Green's function
\begin{eqnarray}
G^{\theta}_{N+M}(x_1',..., x_N';~x_1,..., x_M) = \langle\Omega|T\left(\phi_{\theta}(x_1')\cdots\phi_{\theta}(x_N')\phi_{\theta}(x_1)\cdots\phi_{\theta}(x_M)\right)|\Omega\rangle.
\label{nmncgreenf}
\end{eqnarray}
As mentioned before, the twisted $N+M$-point Green's function is obtained by replacing the commutative fields $\phi_0$ with noncommutative fields $\phi_\theta$ in the time-ordered 
product of fields. Also, the Fourier transform of (\ref{nmncgreenf})  can be obtained by integrating with respect to the measure 
$ \left(\prod_i d^4x_i'\right)\left(\prod_j d^4x_j\right) e^{i\left(\sum_{i\leq N}p_i'\cdot x_i' - \sum_{j\leq M}p_j\cdot x_j\right)}. $

Integration over $x_i$, $x_i'$ gives us $\widetilde{G}_{\theta}^{N+M}(p_1'\cdots , p_N'; p_1 \cdots ,p_M)$. The residue at the poles in all the momenta multiplied together gives 
the scattering amplitude. This is just the noncommutative version of the LSZ reduction formula. We now show that it gives the same expression for the S-matrix elements, as the 
one obtained in previous section using interaction picture.

As done in the commutative case, the pole in $p_1'$ can be obtained by Fourier transforming in just $x_1'$, i.e.
\begin{eqnarray}
\widetilde{G}_{\theta}^{(1)}(p_1',\cdots ,x_N', x_1, \cdots ,x_M) & = & \int d^4x_1' ~e^{i\left(p_1'^0x_1'^0 - \vec{p}_1'\cdot\vec{x}_1'\right)} \, \langle\Omega|
\mathcal{T} \left(\phi_{\theta}(x_1')\cdots\phi_{\theta}(x_N')\phi_{\theta}(x_1) \right. \nonumber \\
 & & \left. \cdots\phi_{\theta}(x_M)\right)|\Omega\rangle.
\end{eqnarray}
Taking $T_+ >> x_N'^0\cdots x_2'^0, x_M^0, \cdots ,x_1^0$, we can isolate the term with pole in $\widetilde{G}_{\theta}^{(1)}$. Hence
\begin{eqnarray}
 \widetilde{G}_{\theta}^{(1)}(p_1',\cdots ,x_N', x_1\cdots x_M) & = & \sqrt{Z}\int_{T_+}^{\infty} dx_1'^0 d^3x_1'~ e^{i\left(p_1'^0x_1'^0 - \vec{p}_1'\cdot\vec{x}_1'\right)} 
\langle\Omega|\phi_{\theta}^{\textrm{out}}(x_1') \mathcal{T} \left(\phi_{\theta}(x_2')\cdots \right.  \nonumber \\
& &  \left. \phi_{\theta}(x_N')\phi_{\theta}(x_1)\cdots\phi_{\theta}(x_M)\right)|\Omega\rangle 
\, + \, \textrm{OT} 
\nonumber \\ 
& = & \sqrt{Z}\int_{T_+}^{\infty}dx_1'^0d^3x_1'\frac{1}{(2\pi)^3} \frac{d^3q_1}{2E_{\vec{q}_1}}e^{i\left(p_1'^0x_1'^0 - \vec{p}_1'\cdot\vec{x}_1'\right)} 
\langle\Omega|\phi_{\theta}^{\textrm{out}}(x_1')|\hat{q}_1\rangle \nonumber \\ 
& &  \langle\hat{q}_1| \mathcal{T} \left(\phi_{\theta}(x_2')\cdots\phi_{\theta}(x_N')\phi_{\theta}(x_1)
\cdots\phi_{\theta}(x_M)\right)|\Omega\rangle  +  \textrm{OT} 
\end{eqnarray}
where
\begin{eqnarray}
{\langle\Omega|\phi_{\theta}^{\textrm{out}}(x_1')|\hat{q}_1\rangle = \langle\Omega|\phi_0^{\textrm{out}}(x_1')|\hat{q}_1\rangle }
\end{eqnarray}
because the twist gives just 1 in this case. This can be seen by using the dressing transformation (\ref{fielddress}), i.e. writing $\phi_{\theta}^{\textrm{out}}$ 
as $e^{\frac{1}{2}\partial \wedge P} \phi_{0}^{\textrm{out}}$ and acting with $P_{\nu}$ on $\langle\Omega|$. 

Repeating essentially the same procedure as in the commutative case, one can extract the pole $\frac{1}{p_1'^2-m^2-i\epsilon}$ and its coefficient. 

For poles at $p'_1$, $p'_2$, we have
 \begin{eqnarray}
& & \widetilde{G}_{\theta}^{(2)}(p_1', p_2', x_3', \cdots , x_N', x_1, \cdots ,x_M) \, = \, \int_{T_+}^{\infty} d^4x_1' d^4x_2'~ e^{ip_1'\cdot x_1' + ip_2'\cdot x_2'} 
(\sqrt{Z})^2\frac{d^3\hat{q}_1d^3\hat{q}_2}{2! (2E_{\vec{q}_1})(2E_{\vec{q}_2})}  \nonumber \\
& & \langle\Omega|\phi_{\theta}^{\textrm{out}}(x_1')\phi_{\theta}^{\textrm{out}}(x_2')|\hat{q}_1,
 \hat{q}_2\rangle\langle\hat{q}_1, \hat{q}_2| \mathcal{T} \left(\phi_{\theta}(x_3')\cdots\phi_{\theta}(x_N')\phi_{\theta}(x_1)\cdots\phi_{\theta}(x_M)\right)|\Omega\rangle  
+ \textrm{OT}.
\end{eqnarray}
 Because of (\ref{id}) there is no twist factor in $|\hat{q}_1, \hat{q}_2\rangle$ and $\langle\hat{q}_2, \hat{q}_1|$. 

Now we compute the matrix element of the two out-fields:
\begin{eqnarray}
\langle\Omega|\phi_{\theta}^{\textrm{out}}(x_1')\phi_{\theta}^{\textrm{out}}(x_2')|\hat{q}_1, \hat{q}_2\rangle  
& = & \int \left(\frac{1}{(2\pi)^3}\right)^2 \frac{d^3p_1''}{E_{\vec{p''}_1}}\frac{d^3p_2''}{2E_{\vec{p''}_2}}e^{-i\hat{p}_1''\cdot x_1' - 
i\hat{p}_2''\cdot x_2'} e^{-\frac{i}{2}\hat{p}_1''\wedge\left(-\hat{p}_2''+\hat{q}_1 + \hat{q}_2\right)} 
\nonumber \\ 
& & e^{-\frac{i}{2}\hat{p}_2''\wedge\left(\hat{q}_1 + \hat{q}_2\right)} \, \langle\Omega|c^{\textrm{out}}_{p_1''}c^{\textrm{out}}_{p_2''}c^{\dag\textrm{out}}_{q_2}
c^{\dag\textrm{out}}_{q_1}|\Omega\rangle,
\label{mel}
\end{eqnarray}
where the matrix element is
\begin{eqnarray}
\langle\Omega|c^{\textrm{out}}_{p_1''}c^{\textrm{out}}_{p_2''}c^{\dag\textrm{out}}_{q_2}c^{\dag\textrm{out}}_{q_1}|\Omega\rangle  
& = & \left(2\pi\right)^3\left(2\pi\right)^3 \,2E_{\vec{p''}_1} 2E_{\vec{p''}_2} \, \left[\delta^3(\vec{p''}_1-\vec{q}_1)\delta^3(\vec{p''}_2-\vec{q}_2) \right. \nonumber \\
& + & \left. \delta^3(\vec{p''}_1-\vec{q}_2)\delta^3(\vec{p''}_2-\vec{q}_1)\right].
\label{momcon}
\end{eqnarray}
It is then clear that the whole matrix element in (\ref{mel}) vanishes unless 
\begin{eqnarray}
\hat{p}_1'' + \hat{p}_2'' = \hat{q}_1 + \hat{q}_2 ,
\label{momcond}
\end{eqnarray}
so that
\begin{eqnarray}
{e^{-\frac{i}{2}\hat{p}_1''\wedge\left(-\hat{p}_2'' + \hat{p}_1'' + \hat{p}_2''\right) -\frac{i}{2}\hat{p}_2''\wedge\left(\hat{p}_1'' + \hat{p}_2''\right)} =
 e^{-\frac{i}{2}\hat{p}_2'' \wedge\hat{p}_1''}.}
\end{eqnarray}

Now, integrations over $\vec{x}_1'$, $\vec{x}_2'$ give us further $\delta$-functions which imply that
\begin{eqnarray}
{\vec{p''}_1=\vec{p'}_1~,~ \vec{p''}_2 = \vec{p'}_2}
\end{eqnarray}
 and hence
\begin{eqnarray}
{\hat{p}_1''=\hat{p}_1'~,~ \hat{p}_2'' = \hat{p}_2'.}
\end{eqnarray}
Thus we finally obtain the noncommutative phase $e^{-\frac{i}{2}\hat{p}_2'\wedge\hat{p}_1'}$. 

Moreover, since
\begin{eqnarray}
{_{\textrm{out}}\langle\hat{q}_1, \hat{q}_2| \rightarrow~ _{\textrm{out}}\langle\hat{p}_1', \hat{p}_2'|}
\end{eqnarray}
 and due to the identity
\begin{eqnarray}
{_{\textrm{out}}\langle\Omega|c^{\textrm{out}}_{q_1}c^{\textrm{out}}_{q_2} = ~_{\textrm{out}}\langle\Omega|c^{\textrm{out}}_{q_2}c^{\textrm{out}}_{q_1}},
\end{eqnarray}
 we finally obtain
\begin{eqnarray}
& & \widetilde{G}_{\theta}^{(2)}(p_1', p_2', \cdots , x_N', x_1, \cdots , x_M) = \frac{\sqrt{Z}}{p_1'^2 - m^2 - i\epsilon}\frac{\sqrt{Z}}{p_2'^2 - m^2 - 
i\epsilon} e^{-\frac{i}{2}\hat{p}_2'\wedge\hat{p}_1'} \nonumber \\ 
& &  _{\textrm{out}}\langle\hat{p}_1' \hat{p}_2'|\mathcal{T}\left(\phi_{\theta}(x_3')\cdots\phi_{\theta}(x_N')\phi_{\theta}(x_1)\cdots\phi_{\theta}(x_M)\right)|\Omega\rangle 
\, + \, \textrm{OT}.
\end{eqnarray}
 The phase can be absorbed so that the twisted out-state becomes
\begin{eqnarray}
{\langle\Omega|a_{\theta}^{\textrm{out}}(\hat{p}_2')a_{\theta}^{\textrm{out}}(\hat{p}_1').}
\end{eqnarray}
Hence the two-particle residue gives us the same expression as obtained in (\ref{ncmsmatrix}). 

As shown in \cite{amilcar} the above analysis can be easily generalized to $N$ outgoing particles. For this purpose it is enough to analyze the phases associated with the outgoing 
fields. Indeed, let us look at
\begin{eqnarray}
\label{25.1}
\langle\Omega|a^{\textrm{out}}_{\hat{p}_1'}a^{\textrm{out}}_{\hat{p}_2'}\cdots a^{\textrm{out}}_{\hat{p}_N'}|\hat{q}_1\cdots\hat{q}_N\rangle
\qquad \text{and} \qquad
\langle \hat{q}_1\cdots\hat{q}_N|a^{\dag}_{\hat{p}_N'}\cdots a^{\dag}_{\hat{p}_1'} |\Omega\rangle.
\end{eqnarray}
 The above two matrix elements have phases related with each other by complex conjugation.  One can easily calculate them by using (\ref{tcom}) and moving the twist of $a_{\hat{p}'}$ 
in the first term to the left and in the second term to the right. This will give the appropriate phase seen in (\ref{ncmsmatrix}).

 One can similarly do a computation for incoming particles as well, where the conjugates of (\ref{25.1}) will appear. Putting all this together, the final answer can
easily be seen to be the same as the one obtained in (\ref{ncmsmatrix}).


\section{Renormalization and $\beta$-function}

In this section, we carry out the renormalization of twisted $\phi^4_{\theta,\ast}$ scalar field theory on the Moyal plane. We argue that the twisted theory is renormalizable,
with the renormalization prescription being similar to that of commutative $\phi^4_{0}$ theory. In particular, we explicitly check the above claim by carrying 
out renormalization to one loop, computing the beta-function upto one loop and analyzing the RG flow of coupling. We show that the twisted-$\beta$ function
is essentially the same as the $\beta$ function of the commutative theory. The case of more general pure matter theories will be considered in the next section. 

In this section, we follow the treatment of \cite{ramond} and \cite{srednicki} for the computations in the commutative $\phi^4_{0}$ theory.


 \subsection{Superficial Degree of Divergence}

We begin by analyzing superficial degree of divergence of a generic Feynman diagram for a $\phi^n_{\theta,\ast}$ scalar field theory in d-dimensions. It is easy to see that the 
criterion for superficial degree of divergence will be the same as that for a generic Feynman diagram for a $\phi^n_{0}$ scalar field theory in d-dimensions. The reason is that 
the noncommutative S-matrix (and Feynman diagrams) differ from their commutative counterparts only by an overall noncommutative phase
which does not contribute to the superficial degree of divergence of a diagram. For a generic noncommutative Feynman diagram (involving only scalars) in d-dimensions 
with E external lines, I internal lines and $V_N$ vertices having N-legs (internal or external) attached to them, the superficial degree of divergence D is
\begin{eqnarray}
 D & = & d - \frac{1}{2}(d - 2)E  + V_N \left( \frac{N-2}{2}d - N \right) .
\label{sdd}
\end{eqnarray}
In $d=4$ dimensions this reduces to
\begin{eqnarray}
 D & = & 4 - E  + V_N \left( N - 4 \right). 
\label{sdd4}
\end{eqnarray}
Furthermore, for  $\phi^4_{\theta,\ast}$ theory in $d=4$ dimensions we have 
\begin{eqnarray}
 D & = & 4 - E  .
\label{ncsdd4}
\end{eqnarray}
We notice that, as expected, the superficial degree of divergences in (\ref{sdd}), (\ref{sdd4}) and (\ref{ncsdd4}) are all the same as that for commutative case. 
So the criterion for determining which of the diagrams will be divergent, remains the same, i.e. the diagrams with $D  \geq 0$ 
are the divergent ones.  Thus, it follows immediately from (\ref{ncsdd4}), that for  $\phi^4_{\theta,\ast}$ theory in $d=4$ dimensions, which is the model we are presently 
interested in, there are divergences for $E=2$ and $E=4$. These correspond to the one particle irreducible (1PI) 2-point function $\Gamma^{(2)}_{\theta}$ and 4-point 
function $\Gamma^{(4)}_{\theta}$ respectively, implying that, $\Gamma^{(2)}_{\theta}$ and $\Gamma^{(4)}_{\theta}$ will be divergent. We need to renormalize them, resulting in 
corrections to propagators and vertices. Furthermore, like in commutative case, by making 1PI two-point function and four-point functions finite, we can make the whole theory finite, 
as these two functions are the only source of divergences.

We further remark that, like in commutative case, just because the superficial degree of divergence of a given diagram is less than zero does not mean that it is divergence free, 
as it can have divergent sub-diagrams. But if we renormalize $\Gamma^{(2)}_{\theta}$ and $\Gamma^{(4)}_{\theta}$, all these sub-divergences 
will be taken into account, resulting in the renormalized theory being divergence free.


\subsection{Dimensional Regularization and Renormalization using the Minimal Subtraction Scheme}

In this section, we carry out the renormalization of $\phi^4_{\theta,\ast}$ scalar field theory on Moyal Plane, using dimensional regularization and minimal subtraction scheme. 
We use $\overline{MS}$ scheme and dimensional regularization by working in d = 4 - $\epsilon$ dimensions. In  d = 4 - $\epsilon$ dimensions the coupling $\lambda$ is no longer 
dimensionless, so we change it to $\lambda \rightarrow \lambda \tilde{\mu}^{\epsilon}$, where $\tilde{\mu}$ is a mass parameter. 

The bare ($\tilde{\phi}_{\theta}$, $m_B$ and $\lambda_B$) and renormalized ($\phi_{\theta}$, $m$ and $\lambda$) fields and parameters are related with each other as 
\begin{eqnarray}
\tilde{\phi}_{\theta} & = & Z^{1/2}_{\phi} \, \phi_{\theta} , \nonumber \\
m_B & = & Z^{-1/2}_{\phi} \,  Z_{m} \, m , \nonumber \\
\lambda_B & = & Z^{-2}_{\phi} \, Z_{\lambda} \, \lambda \, \tilde{\mu}^{\epsilon} .
\label{b2r}
\end{eqnarray}
where $Z_{\Phi}$ is the wavefunction renormalization constant, $Z_m$ is the mass renormalization constant and $Z_{\lambda}$ is the coupling renormalization constant. 
The Zs are as of yet unknown constants and are to be evaluated perturbatively. It should also be noted that the functional form of the Zs depends on the renormalization scheme. 
Moreover, it turns out that in $\overline{MS}$ renormalization scheme, the Zs will have a generic form like
\begin{eqnarray}
 Z_{\Phi} & = & 1 \, + \, \sum_{n=1}^{\infty}\, \frac{a_n(\lambda)}{\epsilon^n}, \nonumber \\
 Z_{m} & = & 1 \, + \, \sum_{n=1}^{\infty}\, \frac{b_n(\lambda)}{\epsilon^n}, \nonumber \\
 Z_{\lambda} & = & 1 \, + \, \sum_{n=1}^{\infty}\, \frac{c_n(\lambda)}{\epsilon^n}.
\label{zstrt}
\end{eqnarray}
From (\ref{zstrt}) and as we will argue later in this section, the Zs are all independent of $\theta$ to all orders in perturbation theory. This implies that the $\beta$-function, 
the anomalous dimensions of mass and n-point Green's functions will be the same as that for commutative $\phi^4_0$ theory.


\subsection{2-Point Function}

The Feynman diagrams contributing at one loop to the two-point function are seen in figure 1.

\begin{figure}[h!]
        \centerline{
               \mbox{\includegraphics*[angle=0,width=7in]{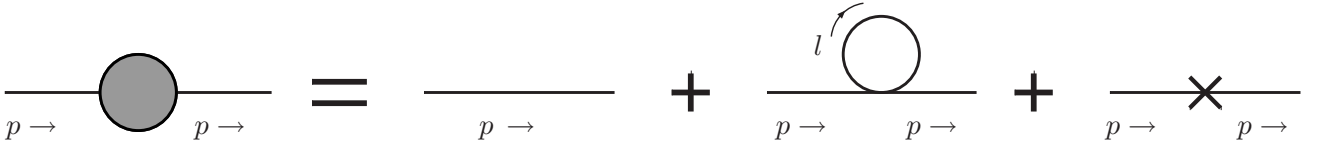}}
               }
 \caption{ Feynman diagrams for the 2-point function at one loop.  }
               \end{figure}
               
So the loop contribution to the 2-point function is given by
\begin{eqnarray}
 -i \Pi(k^2) & = &  \frac{1}{2}(-i Z_{ \lambda}\,  \lambda \, \tilde{\mu}^{\epsilon}) \int \frac{d^d l}{(2\pi)^d}\, \frac{i}{l^2 - m^2} \, + \, i (A \, k^2 - Bm^2),
\label{twopointfunction}
\end{eqnarray}
where $A = Z_{\phi} - 1$ and $B = Z_m - 1$.

Now, let us consider the integral
\begin{eqnarray}
\xi & = &  \tilde{\mu}^{\epsilon} \int \frac{d^d l}{(2\pi)^d}\, \frac{i}{l^2 - m^2}.
\label{twopointintegral}
\end{eqnarray}
Substituting $l^0 = i l^0_E$ and going to Euclidean plane we have 
\begin{eqnarray}
\xi & = &  \tilde{\mu}^{\epsilon} \int \frac{d^d l_E}{(2\pi)^d}\, \frac{1}{l_E^2 + m^2},
\label{2pointintegral}
\end{eqnarray}
where $l_E^2 = (l_E^0)^2 + \vec{l}^2 $. The integral evaluates to (d = 4 - $\epsilon$) \cite{srednicki}
\begin{eqnarray}
 \xi & = & \frac{\Gamma(-1 + \frac{\epsilon}{2})}{(4\pi)^2} \, m^2 \, \left( \frac{4\pi\tilde{\mu}^2}{m^2} \right)^{\frac{\epsilon}{2}}
\label{2pointinteg}
\end{eqnarray}
Now, we use the identity
\begin{eqnarray}
 \Gamma (-n + x) & = & \frac{(-1)^n}{n!} \left[ \frac{1}{x} - \gamma + \sum^n_{k=1} k^{-1} + O(x)\right],
\label{gammaidentity}
\end{eqnarray}
where $\gamma$ is the Euler-Mascheroni constant. Using (\ref{gammaidentity}) in (\ref{2pointinteg}) we obtain
\begin{eqnarray}
\xi & = & \frac{-m^2}{(4\pi)^2} \, \left( \frac{2}{\epsilon} - \gamma + 1 \right) \, \left( \frac{4\pi\tilde{\mu}^2}{m^2} \right)^{\frac{\epsilon}{2}} \nonumber \\
& = & \frac{-m^2}{(4\pi)^2} \, \left[ \left( \frac{2}{\epsilon} - \gamma + 1 \right) \, + \,\ln \left( \frac{4\pi\tilde{\mu}^2}{m^2} \right) \, + \, (- \gamma + 1) \frac{\epsilon}{2}
 \ln \left( \frac{4\pi\tilde{\mu}^2}{m^2}\right) \right],
\label{2pointint}
\end{eqnarray}
where we have used the relation $X^{\frac{\epsilon}{2}} = 1 + \frac{\epsilon}{2} \ln X + O(\epsilon^2),\, \text{for} \, \epsilon << 1$. Since we are interested 
in the d = 4 case, we take the limit $\epsilon \rightarrow 0$ in (\ref{2pointint}), so that 
\begin{eqnarray}
\lim_{\epsilon \rightarrow 0} \xi & = & \frac{-m^2}{(4\pi)^2} \, \left[  \frac{2}{\epsilon}  + 1 \, + \,\ln \left(\frac{\mu^2}{m^2} \right) \right],
\label{lim2pointint}
\end{eqnarray}
where we have $\mu^2 = 4\pi\tilde{\mu}^2 \, e^{-\gamma}$. Using (\ref{lim2pointint}) in (\ref{twopointfunction}) we obtain
\begin{eqnarray}
 \lim_{\epsilon \rightarrow 0} -i \Pi(k^2) & = &  \frac{(-i  \lambda)}{2} \, \frac{-m^2}{(4\pi)^2} \, \left[  \frac{2}{\epsilon}  + 1 \, + \,\ln \left(\frac{\mu^2}{m^2} \right) \right]
 \, + \, i (A \, k^2 - B \, m^2) .
\label{limtwopointfunction}
 \end{eqnarray}
As can be seen from (\ref{limtwopointfunction}), the singularities due to loop contribution manifest themselves as certain terms developing singularities in the 
limit $\epsilon \rightarrow 0$. Since we are interested in only the singular terms we may split (\ref{limtwopointfunction}) as
\begin{eqnarray}
 \lim_{\epsilon \rightarrow 0} \Pi(k^2) & = & - \frac{ \lambda \, m^2}{(4\pi)^2} \, \frac{1}{\epsilon} \, - \, A \, k^2  +  B \, m^2 \, + \, \text{Terms of finite order}.
\label{limtwopointfn}
\end{eqnarray}

Now, according to the $\overline{MS}$ scheme, the constants $A$ and B are to be chosen in such a way as to cancel all the singular terms in (\ref{limtwopointfn}). So we have
\begin{eqnarray}
  A & = & Z_{\phi} - 1 \; = \; O( \lambda^2) \qquad \qquad \qquad \; \; \; \;  \Rightarrow \, Z_{\phi} \; = \; 1 \, + \, O( \lambda^2) \nonumber \\
B & = & Z_{m} - 1 \; = \; \frac{ \lambda}{16 \pi^2} \, \frac{1}{\epsilon} \, + \, O( \lambda^2 ) \qquad \Rightarrow \, Z_{m} \; = \; 1 \, + \,  \frac{ \lambda}{16 \pi^2}
 \, \frac{1}{\epsilon} \, + \, O( \lambda^2 )
\label{ab}
\end{eqnarray}

\subsection{4-Point Function}

The Feynman diagrams up to one loop for the four-point function are depicted in figure 2.

\begin{figure}[H]
        \centerline{
               \mbox{\includegraphics*[angle=0,width=7in]{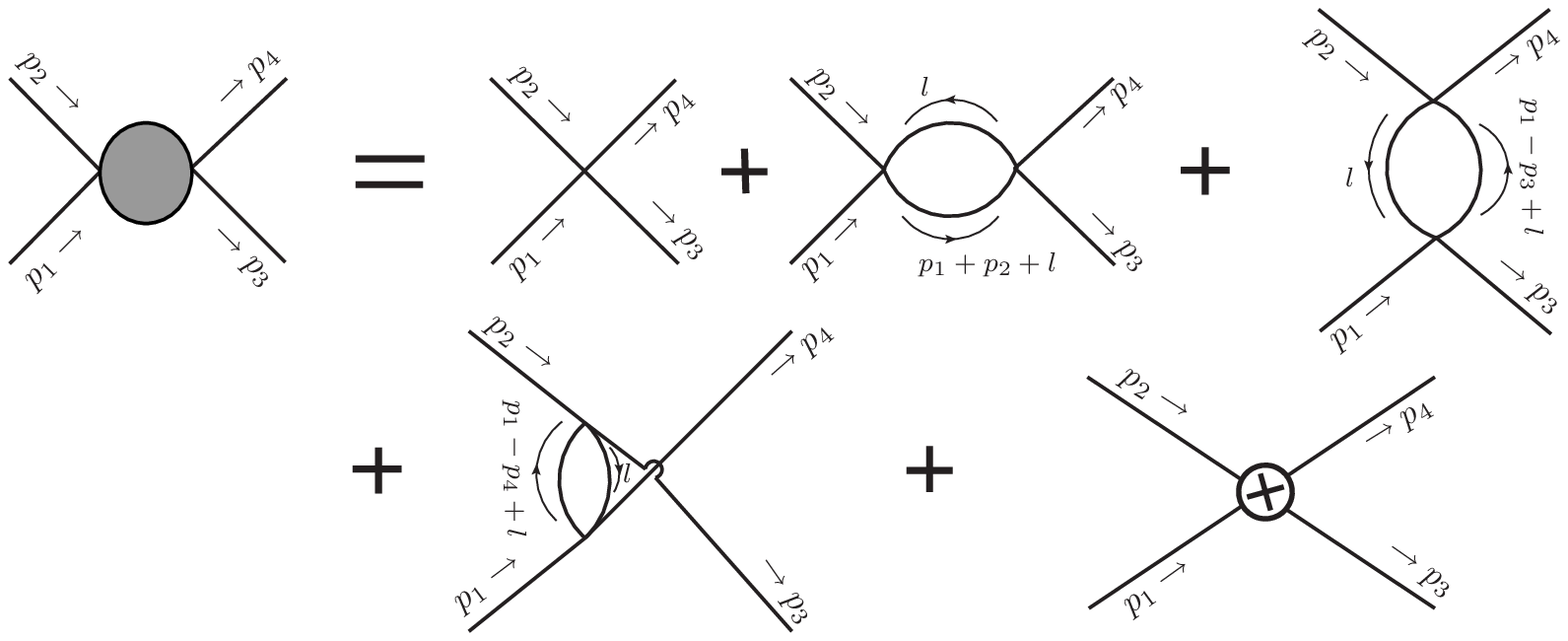}}
               }
 \caption{ Feynman diagrams for the 4-point function at one loop. }
               \end{figure}

The 4-point function is given by
\begin{eqnarray}
i \Gamma^{(4)}_{\theta}  =  e^{\frac{i}{2}\left(p_1\wedge p_2 - p_3\wedge p_4\right)}  \left[-i Z_{\lambda} \lambda  \tilde{\mu}^{\epsilon}  
  + \frac{1}{2} \left( -i Z_{\lambda} \lambda  \tilde{\mu}^{\epsilon} \right)^2  \left \{ i V(s) + iV(t) + iV(u) \right \}  +  O(\lambda^3) \right]
\label{fourpointfunction}
\end{eqnarray}
where s, t, u are the Mandelstam variables defined as $ s = (p_1 + p_2)^2$, $ t = (p_3 - p_1)^2$ and $ u = (p_4 - p_1)^2$ and 
\begin{eqnarray}
 iV(p^2) & = & \int \frac{d^d l}{(2\pi)^d} \, \frac{i}{(l + p)^2 - m^2} \, \frac{i}{l^2 - m^2}.
\label{4pointintegral}
\end{eqnarray}
The appearance of noncommutative phases in (\ref{fourpointfunction}) is an attribute of the twisted statistics followed by the particles. Moreover, these phases insure that 
$\Gamma^{(4)}_{\theta} $ has right symmetries vis-a-vis twisted Poincar\'e invariance.

Now, consider the integral
\begin{eqnarray}
\tilde{\mu}^{\epsilon} \, iV(p^2) & = &  \tilde{\mu}^{\epsilon} \, \int \frac{d^d l}{(2\pi)^d} \, \frac{i}{(l + p)^2 - m^2} \, \frac{i}{l^2 - m^2},
\label{div4pointintegral}
\end{eqnarray}
which evaluates after Wick rotation $q^0 \rightarrow iq^0_E$ to \cite{srednicki}
\begin{eqnarray}
 \tilde{\mu}^{\epsilon} \, iV(p^2) & = &  - i \tilde{\mu}^{\epsilon} \, \int_0^1 dx \, \int \frac{ d^d q_E}{(2\pi)^d} \, \frac{1}{[q_E^2 + D]^2}
\label{d4ptintegral}
\end{eqnarray}
where $D = m^2 - x(1-x)p^2 $, with $x$ being a Feynman parameter. 

Using the standard integral
\begin{eqnarray}
 \int \frac{ d^d q_E}{(2\pi)^d} \, \frac{(q_E^2)^a}{[q_E^2 + D]^b} & = & \frac{\Gamma(b - a - \frac{d}{2}) \,
 \Gamma(a + \frac{d}{2})}{(4\pi)^{\frac{d}{2}}\,\Gamma(b)\,\Gamma(\frac{d}{2})},
\, D^{-(b - a - \frac{d}{2})}
\label{stdint}
\end{eqnarray}
we have
\begin{eqnarray}
 \tilde{\mu}^{\epsilon} \, iV(p^2) & = &  - i \tilde{\mu}^{\epsilon} \, \int_0^1 dx \, \frac{\Gamma(2 - \frac{d}{2})}{(4\pi)^{\frac{d}{2}}} \, D^{-(2 - \frac{d}{2})} .
\label{d4pintegral}
\end{eqnarray}
Putting $ d = 4 - \epsilon $, we have
\begin{eqnarray}
 \tilde{\mu}^{\epsilon} \, iV(p^2) 
& = &  \frac{-i}{(4\pi)^2} \, \Gamma \left( \frac{ \epsilon}{2} \right ) \, \int_0^1 dx \, \left( \frac{4\pi \, \tilde{\mu}^{2}}{D} \right )^{\frac{ \epsilon}{2}}.
\label{d4pinteg}
\end{eqnarray}
Using the identity
\begin{eqnarray}
 \Gamma(-n + x) & = & \frac{(-1)^n}{n!}\, \left[ \frac{1}{x} - \gamma + \sum_{k=1}^n \, k^{-1} + O(x) \right],
\label{gammafn}
\end{eqnarray}
we have 
\begin{eqnarray}
 \Gamma \left( \frac{ \epsilon}{2} \right ) & = &  \frac{2}{\epsilon} - \gamma +  O(\epsilon) .
\label{gamma}
\end{eqnarray}
Using (\ref{gamma}) in (\ref{d4pinteg}) we have
\begin{eqnarray}
 \tilde{\mu}^{\epsilon} \, iV(p^2) & = & \frac{-i}{(4\pi)^2} \, \left( \frac{2 }{\epsilon} - \gamma \right) \, \int_0^1 dx \, 
\left( \frac{4\pi \, \tilde{\mu}^{2}}{D} \right )^{\frac{ \epsilon}{2}}.
\label{d4pinte}
\end{eqnarray}
In the limit $\epsilon \rightarrow 0$, we have
\begin{eqnarray}
 \lim_{\epsilon \rightarrow 0} \tilde{\mu}^{\epsilon} \, iV(p^2) & = &  \frac{-i}{(4\pi)^2} \, \left[ \frac{2 }{\epsilon} +   \int_0^1 dx \, \ln \left( \frac{\mu^{2}}{ D} \right) \right],
\label{limd4pint}
\end{eqnarray}
Using (\ref{limd4pint}) into (\ref{fourpointfunction}) we obtain
\begin{eqnarray}
\lim_{\epsilon \rightarrow 0} i \Gamma^{(4)}_{\theta} & = & e^{\frac{i}{2}\left(p_1\wedge p_2 - p_3\wedge p_4\right)} \, \left[-i Z_{\lambda}\, \lambda \, 
 \, + \, \frac{1}{2} \left( -i Z_{\lambda}\, \lambda \right)^2 \, \left( \frac{-i}{(4\pi)^2} \right) \, \left \{ \frac{6 }{\epsilon} +  \int_0^1 dx 
\, \left(  \ln \left( \frac{\mu^{2}}{ D(s)} \right)  \right. \right. \right.  \nonumber \\
& + & \left.\left. \left. \ln \left( \frac{\mu^{2}}{ D(t)} \right)  \, + \, \ln \left( \frac{\mu^{2}}{ D(u)} \right) \right) \right \} + O(\lambda^3) \right]  \nonumber \\
& \approx & e^{\frac{i}{2}\left(p_1\wedge p_2 - p_3\wedge p_4\right)} \left[-i Z_{\lambda} \lambda  
  +  \left( -i \lambda \right)^2 \left( \frac{-i}{32\pi^2} \right)  \left \{ \frac{6 }{\epsilon} +  \text{Finite Terms} \right \} + O(\lambda^3) \right] 
\label{4pointfunction}
\end{eqnarray}
where in writing last line we have neglected higher powers of $Z_{\lambda}$.

Now, in accordance with $\overline{MS}$ scheme, matching the divergent parts in (\ref{4pointfunction}), we obtain
\begin{eqnarray}
 Z_{\lambda} & = & 1 + \frac{3\lambda}{16 \pi^2}\, \frac{1}{\epsilon} 
\label{zlambda}
\end{eqnarray}
which is the same as that for the commutative theory. Note that, as remarked in the beginning of this section, the $Z_{\lambda}$, $Z_{\phi}$ and $Z_{m}$ are all completely 
independent of $\theta$. This is what we naively expected from our analysis of the tree level theory in previous sections. The noncommutative corrections are just phases. 
Hence they do not result in any new source of divergence. Moreover, since the form of Zs is completely fixed (within a given renormalization scheme) by the demand that the 
renormalized theory 
should be divergence free, if we try to put an implicit dependence of $\theta$ in Zs, then (\ref{zlambda}) and (\ref{ab}) will not be satisfied, implying that the renormalized 
theory is still not completely free from divergences. So the demand that renormalized theory be completely free of any divergence, forces us to choose Zs of the form (\ref{zlambda}) 
and (\ref{ab}) and hence no dependence of Zs on $\theta$, whether implicit or explicit, is allowed. Moreover, although we have done calculations with a particular 
renormalization scheme, it is easy to see that whatever renormalization scheme one chooses to use, the source and form of divergences always remains the same. The noncommutative 
phases will never result in any new divergence or contribute to any divergence and hence the demand to cancel all the divergences will always imply that at least the divergent 
part of Zs is completely independent of $\theta$. As it does in commutative theory, the prescription dependence of renormalization scheme will only effect 
the finite terms. Hence, even changing renormalization scheme or for that matter, even the regularization technique, does not change the essential result that the divergent part 
of Zs have no dependence, implicit or explicit, on $\theta$. 

\textbf{Higher Loop Corrections to 2-Point and 4-Point Functions :} 

Although in this paper we restrict ourself only to one loop corrections to 2-point and 4-point functions, higher loop effects can similarly be computed. 
The noncommutative correction are always a phase (to all orders of perturbation). They never give rise to new sources of divergences. So the Zs to all orders in perturbation will 
be always independent (implicitly as well as explicitly) of $\theta$ and will have the same form as that of commutative theory. So, like in commutative case the generic form of Zs 
are   
\begin{eqnarray}
 Z_{\phi} & = & 1 \, + \, \sum_{n=1}^{\infty}\, \frac{a_n(\lambda)}{\epsilon^n} \nonumber \\
 Z_{m} & = & 1 \, + \, \sum_{n=1}^{\infty}\, \frac{b_n(\lambda)}{\epsilon^n} \nonumber \\
 Z_{\lambda} & = & 1 \, + \, \sum_{n=1}^{\infty}\, \frac{c_n(\lambda)}{\epsilon^n}
\label{1zstrt}
\end{eqnarray}
where $a_n(\lambda)$, $b_n(\lambda)$ and $c_n(\lambda)$ are unknown functions which are evaluated perturbatively by demanding that the renormalized theory be independent of 
divergences at all orders of perturbation. Also note that $a_n(\lambda)$, $b_n(\lambda)$ and $c_n(\lambda)$ are all independent of $\theta$ and as argued before they have the 
same form as for 
commutative $\phi^4_0$ theory.


\subsection{Renormalization Group and $\beta$-Function}

In previous section we showed that all Zs are independent of $\theta$ and are the same as in the commutative case. In view of this, we expect and will show by explicit computations 
that it is indeed the case. The $\beta$-function and R.G equation are also independent of $\theta$. They are the same as in the commutative case.

For $\beta$-function computation we start with noticing the fact that the bare and renormalized couplings are related with each other via
\begin{eqnarray}
 \lambda_B & = & Z^{-2}_{\phi} \, Z_{\lambda} \, \lambda \, \tilde{\mu}^{\epsilon},
\label{btor}
\end{eqnarray}
or
\begin{eqnarray}
\ln \lambda_B & = & \ln (Z^{-2}_{\phi} \, Z_{\lambda}) \, + \, \ln \lambda \, + \, \epsilon \, \ln \tilde{\mu}
\label{lnbtor}
\end{eqnarray}
Differentiating (\ref{lnbtor}) with respect to $\ln \mu$, we obtain
\begin{eqnarray}
\frac{\partial (\ln \lambda_B)}{\partial(\ln \mu)} & = & \frac{\partial (\ln (Z^{-2}_{\phi} \, Z_{\lambda}))}{\partial (\ln \mu)} \, + \, \frac{\partial(\ln \lambda)}{\partial (\ln \mu)}
 \, + \, \frac{\partial (\epsilon \, \ln \tilde{\mu})}{\partial (\ln \mu)} \nonumber \\
& = & \frac{\partial (\ln (Z^{-2}_{\phi} \, Z_{\lambda}))}{\partial (\ln \mu)} \, + \, \frac{\partial (\ln \lambda)}{\partial (\ln \mu)}
 \, + \, \epsilon .
\label{dlnbtor}
\end{eqnarray}
where $\mu$ is a mass scale, $\mu^{2} = 4\pi \, e^{-\gamma}\, \tilde{\mu}^{2} $. 
Now, we demand that the bare coupling be independent of $\mu$, i.e. $\frac{\partial (\ln \lambda_B)}{\partial (\ln \mu)} = 0 $. Then
\begin{eqnarray}
 0 & = & \frac{\partial (\ln (Z^{-2}_{\phi} \, Z_{\lambda}))}{\partial (\ln \mu)} \, + \, \frac{\partial (\ln \lambda)}{\partial (\ln \mu)}
 \, + \, \epsilon \nonumber \\
& = & \frac{\partial (\ln (Z^{-2}_{\phi} \, Z_{\lambda}))}{\partial\, \lambda} \, \frac{\partial\, \lambda}{\partial (\ln \mu)} 
\, + \, \frac{1}{\lambda} \, \frac{\partial \, \lambda}{\partial (\ln \mu)} \, + \, \epsilon .
\label{dlnbtorb0}
\end{eqnarray}
From (\ref{ab}) and (\ref{zlambda}) we have 
\begin{eqnarray}
 \ln (Z^{-2}_{\phi} \, Z_{\lambda}) & = & \ln \left( 1 + \frac{3\lambda}{16 \pi^2}\, \frac{1}{\epsilon} \right) \; = \; \frac{3\lambda}{16 \pi^2}\, \frac{1}{\epsilon} 
\, + \, O(\lambda^2) .
\label{lnzform}
\end{eqnarray}
Using (\ref{lnzform}) in (\ref{dlnbtorb0}) we obtain 
\begin{eqnarray}
\frac{3}{16 \pi^2}\,\frac{1}{\epsilon}\,\frac{\partial\,\lambda}{\partial (\ln \mu)} \, + \, \frac{1}{\lambda}\, \frac{\partial\, \lambda}{\partial (\ln \mu)}\, + \, \epsilon & = & 0,
\label{gbetafn1}
\end{eqnarray}
or
\begin{eqnarray}
\frac{\partial\,\lambda}{\partial (\ln \mu)} & = & - \epsilon \lambda \, + \, \frac{3 \lambda^2}{16 \pi^2} \, + \, O(\lambda^3) .
\label{gbetafn}
\end{eqnarray}
Therefore the $\beta$-function is given by
\begin{eqnarray}
 \beta (\lambda) & = & \lim_{\epsilon \rightarrow 0} \, \frac{\partial\,\lambda}{\partial (\ln \mu)} \; = \; \frac{3 \lambda^2}{16 \pi^2} \, + \, O(\lambda^3). 
\label{beta} 
\end{eqnarray}
We note that, as expected, (\ref{beta}) is completely independent of $\theta$ and is the same as the commutative $\beta$-function.

Integrating (\ref{beta}) we can immediately calculate the running of coupling constant with respect to variation in the scale $\mu$, which turns out to be the same as in the 
commutative theory. It is given by
\begin{eqnarray}
 \lambda_2  & = & \frac{\lambda_1}{1 \, - \,\frac{3\lambda_1 }{16 \pi^2} \, \ln \left( \frac{\mu_2}{\mu_1} \right) }.
\label{runningcoupling}
\end{eqnarray}

Now we calculate the R.G equation for a generic n-point 1PI function $\Gamma^{(n)}_{\theta}$. 

The bare n-point 1PI functions $\Gamma^{(n)}_{\theta, B}$ and renormalized n-point 1PI functions  $\Gamma^{(n)}_{\theta, R}$ are related with each other as
\begin{eqnarray}
 \Gamma^{(n)}_{\theta, B} \, (p_1, \dots, p_n, \theta; \, \lambda_B, m_B, \epsilon) & = &  Z^{\frac{-n}{2}}_{\phi} 
\, \Gamma^{(n)}_{\theta, R} \, (p_1, \dots, p_n, \theta; \, \lambda, m, \epsilon, \mu)
\label{n1pifn}
\end{eqnarray}
where all the $\Gamma^{(n)}_{\theta, R}$ are finite as $\epsilon \rightarrow 0$. From (\ref{n1pifn}) we see that the left hand side does not depend on the arbitrary scale $\mu$ 
but the right hand side has explicit as well as implicit (through the mass $m$ and coupling $\lambda$) dependence on $\mu$ \footnote{ Its worth noting that the functional 
dependence of  both $\Gamma^{(n)}_{\theta, B}$ and  $\Gamma^{(n)}_{\theta, R}$ on the noncommutative phases (like on momenta) is same, so the noncommutative phases will not 
affect the R.G. equations}.  So if (\ref{n1pifn}) is correct then the explicit and implicit dependence of $\Gamma^{(n)}_{\theta, R}$ on $\mu$ should cancel each other, i.e.
\begin{eqnarray}
 \frac{ \partial \ln \left \{\Gamma^{(n)}_{\theta, B} \, (p_1, \dots, p_n, \theta; \, \lambda_B, m_B, \epsilon) \right \}}{ \partial \ln \mu}  \, = \, 
\frac{ \partial \ln \left \{ Z^{\frac{-n}{2}}_{\phi} \, \Gamma^{(n)}_{\theta, R} \, (p_1, \dots, p_n, \theta; \, \lambda, m, \epsilon, \mu) \right \}}{ \partial \ln \mu} & = & 0 
\nonumber \\
\left[ \mu \, \frac{\partial }{\partial \mu} \, + \, \mu \, \frac{\partial \lambda}{\partial \mu} \, \frac{\partial}{\partial \lambda} 
\, + \,\mu \, \frac{\partial m }{\partial  \mu} \, \frac{\partial}{\partial m} \, - \, \mu \, \frac{n}{2} \, \frac{\partial \ln Z_{\phi}}{\partial \mu}  \right]
\, \Gamma^{(n)}_{\theta, R} \, (p_1, \dots, p_n, \theta; \, \lambda, m, \epsilon, \mu) = 0 \nonumber \\
\left[ \mu \, \frac{\partial }{\partial \mu} \, + \, \beta \, \frac{\partial}{\partial \lambda} \, + \, \frac{1}{m} \,\gamma_m \, \frac{\partial}{\partial m} \, - \, n \gamma_d  \right]
\, \Gamma^{(n)}_{\theta, R} \, (p_1, \dots, p_n, \theta; \, \lambda, m, \epsilon, \mu) = 0
\label{nrg}
\end{eqnarray}
where $\beta$ is the $\beta$-function, $\gamma_m$ is the anomalous mass dimension and $\gamma_d$ is the anomalous scaling dimension of $\Gamma^{(n)}_{\theta, R}$.

The equations (\ref{nrg}) are the R.G equations for a noncommutative n-point 1PI function. The noncommutativity does not give rise to any new divergences. Since the functional 
dependence of bare and renormalized n-point 1PI functions on noncommutative parameters are same, the R.G equations are essentially the same as that for commutative theory. 
Hence the noncommutative phases in (\ref{nrg}) sit more like spectators and do not affect the R.G. equations.


\section{Generic Pure Matter Theories}

So far we have restricted ourselves to the case of noncommutative real scalar fields having a $\phi^4_{\theta,\ast}$ self interaction. In this section, we consider more general 
noncommutative theories with polynomial interactions and involving only matter fields. As we show in this section, the formalism developed and discussed in previous sections, for
real scalar fields, goes through (with appropriate generalizations) for all such theories. Noncommutative theories involving gauge fields need a separate treatment and will 
not be discussed in this work.

\subsection{Complex Scalar Fields}
 
Let $\phi_{\theta}$ be a noncommutative complex scalar field having a normal mode expansion
\begin{eqnarray}
 \phi_{\theta} (x) & = & \int d^3 \tilde{k} \left[ a_k \, e^{-ikx} \, + \, b^\dagger_k e^{ikx}  \right] \nonumber \\
\phi^\dagger_{\theta} (x) & = & \int d^3 \tilde{k} \left[ b_k \, e^{-ikx} \, + \, a^\dagger_k e^{ikx}  \right] 
\label{csncfields}
\end{eqnarray}
where $ d^3 \tilde{k} \, = \, \frac{d^3 k}{(2\pi)^3 2 E_k}$, and $a_k$, $b_k$ are noncommutative annihilation operators satisfying the twisted algebra:
\begin{eqnarray}
  a^\#_{p_{1}} a^\#_{p_{2}} & = & \eta \, e^{ip_{1}\wedge p_{2}} \,a^\#_{p_{2}} a^\#_{p_{1}} \nonumber \\
(a^{\dagger}_{p_{1}})^\# (a^{\dagger}_{p_{2}})^\# & = & \eta \, e^{ip_{1}\wedge p_{2}} \, (a^{\dagger}_{p_{2}})^\# (a^{\dagger}_{p_{1}})^\# \nonumber \\
a_{p_{1}} a^{\dagger}_{p_{2}} & = & \eta \, e^{-ip_{1}\wedge p_{2}} \, a^{\dagger}_{p_{2}} a_{p_{1}} + (2\pi)^3 \, 2 E_p \,\delta^{3}(p_{1} \, - \, p_{2}) \nonumber \\
b_{p_{1}} b^{\dagger}_{p_{2}} & = & \eta \, e^{-ip_{1}\wedge p_{2}} \, b^{\dagger}_{p_{2}} b_{p_{1}} + (2\pi)^3 \, 2 E_p \,\delta^{3}(p_{1} \, - \, p_{2}) \nonumber \\
a_{p_{1}} b^{\dagger}_{p_{2}} & = & \eta \, e^{-ip_{1}\wedge p_{2}} \, b^{\dagger}_{p_{2}} a_{p_{1}}  
\label{cstcom}
\end{eqnarray}
where $a^\#_{p}$ and $(a^{\dagger}_{p})^\# $ stands for either of the operators $a_p, b_p$ and $a^\dagger_p, b^\dagger_p$ respectively. For $\eta = 1$, these are bosonic operators. 
For $\eta = -1$, these are fermionic operators. We consider in the following $\eta = 1$. 

The  noncommutative creation/annihilation operators are related with their commutative counterparts (denoted by $c_k$, $d_k$ respectively) by the dressing transformations
\begin{eqnarray}
 a_k & = & c_k \, e^{-\frac{i}{2} \, k \wedge P}, \nonumber \\
b_k & = & d_k \, e^{-\frac{i}{2} \, k \wedge P}, \nonumber \\
 a^\dagger_k & = & c^\dagger_k \, e^{\frac{i}{2} \, k \wedge P}, \nonumber \\
b^\dagger_k & = & d^\dagger_k \, e^{\frac{i}{2} \, k \wedge P} 
\label{csdressingtransformations}
\end{eqnarray}
where $P_\mu$ is the Fock space momentum operator
\begin{eqnarray}
 P_\mu & = & \int d^3 \tilde{p} \, p_\mu [c^\dagger_p c_p \, + \, d^\dagger_p d_p] \nonumber \\
& = & \int d^3 \tilde{p} \, p_\mu [a^\dagger_p a_p \, + \, b^\dagger_p b_p]
\label{csmomentumoperator}
\end{eqnarray}
Using (\ref{csncfields}) and (\ref{csdressingtransformations}) one can easily check that the noncommutative fields are also related with commutative fields by the dressing 
transformation
\begin{eqnarray}
 \phi_{\theta} (x) & = & \phi_{0} (x) \, e^{\frac{1}{2} \, \overleftarrow{\partial} \wedge P} \nonumber \\
 \phi^\dagger_{\theta} (x) & = & \phi^\dagger_{0} (x) \, e^{\frac{1}{2} \, \overleftarrow{\partial} \wedge P}
\label{csfdress}
\end{eqnarray}
 where $\phi_{0}$ and $\phi^\dagger_{0}$ are the commutative complex scalar fields having the mode expansion
\begin{eqnarray}
 \phi_{0} (x) & = & \int d^3 \tilde{k} \left[ c_k \, e^{-ikx} \, + \, d^\dagger_k e^{ikx}  \right] \nonumber \\
\phi^\dagger_{0} (x) & = & \int d^3 \tilde{k} \left[ d_k \, e^{-ikx} \, + \, c^\dagger_k e^{ikx}  \right] 
\label{cscomfields}
\end{eqnarray}

Since $\phi_\theta$ is composed of the operators $a_p$ and $b_p$ following twisted statistics, unlike commutative fields, the commutator of $\phi_\theta$ 
and $\phi^\dagger_\theta$ evaluated at same spacetime points does not vanish, i.e.
\begin{eqnarray}
 [\phi_\theta (x), \phi^\dagger_\theta (x)] & \neq & 0
\label{op}
\end{eqnarray}

In view of (\ref{op}), one can in principle write six different quartic self interaction terms which naively seem inequivalent to each other. Hence, a generic 
interaction hamiltonian density with quartic self interactions can be written as
\begin{eqnarray}
& &  \mathcal{H}^{\theta}_{\rm{Int}} (x)  \, = \, \frac{\lambda_1}{4} \, \phi^\dagger_{\theta} \ast \phi^\dagger_{\theta} \ast \phi_{\theta} \ast \phi_{\theta} (x) 
\, + \, \frac{\lambda_2}{4} \,\phi^\dagger_{\theta} \ast \phi_{\theta} \ast \phi^\dagger_{\theta} \ast \phi_{\theta} (x)
  \, + \,\frac{\lambda_3}{4} \, \phi^\dagger_{\theta} \ast \phi_{\theta} \ast \phi_{\theta} \ast \phi^\dagger_{\theta} (x)  \nonumber \\
 & & + \, \frac{\lambda_4}{4} \, \phi_{\theta} \ast \phi^\dagger_{\theta} \ast \phi^\dagger_{\theta} \ast \phi_{\theta} (x)
\, + \, \frac{\lambda_5}{4} \,\phi_{\theta} \ast \phi^\dagger_{\theta} \ast \phi_{\theta} \ast \phi^\dagger_{\theta} (x)
 \, + \,\frac{\lambda_6}{4} \,\phi_{\theta} \ast \phi_{\theta} \ast \phi^\dagger_{\theta} \ast \phi^\dagger_{\theta} (x),
\label{csnchamiltonian}
\end{eqnarray}
where the $\lambda_i$ are the six coupling constants and in general they need not be equal to each other.

\subsection*{Some Identities}

We now list some identities that the noncommutative fields satisfy.
\begin{eqnarray}
\textbf{ 1)} \qquad  &  \phi^\dagger_\theta (x) \, \ast \, \phi_\theta (x)  \, = \,  \phi_\theta (x) \, \ast \, \phi^\dagger_\theta (x)  & \qquad \qquad
\qquad \qquad \qquad \qquad \qquad \qquad \qquad \qquad 
\label{identity1}
\end{eqnarray}

\textbf{Proof : } Using (\ref{csncfields}) we have
\begin{eqnarray}
 \phi^\dagger_\theta (x) \, \ast \, \phi_\theta (x) & = &  \int d^3 \tilde{k}_1 \left[ b_{k_1} \, e^{-ik_1x} \, + \, a^\dagger_{k_1} e^{ik_1x} \right] 
\, e^{\frac{i}{2} \,\overleftarrow{\partial} \wedge \overrightarrow{\partial}}  \, \int d^3 \tilde{k}_2 \left[ a_{k_2} \, e^{-ik_2x} \, + \, b^\dagger_{k_2} e^{ik_2x}  \right]
\nonumber \\
& = & \int d^3 \tilde{k}_1 \, d^3 \tilde{k}_2 \,\left[ b_{k_1} \, e^{-ik_1x} \, e^{\frac{i}{2} (-ik_1) \wedge (-ik_2)} \, a_{k_2} \, e^{-ik_2x}
 \, + \, b_{k_1} \, e^{-ik_1x} \, e^{\frac{i}{2} (-ik_1) \wedge (ik_2)}  \right. \nonumber \\
& + & \left.  b^\dagger_{k_2} e^{ik_2x} a^\dagger_{k_1} e^{ik_1x}  e^{\frac{i}{2} (ik_1) \wedge (-ik_2)}  a_{k_2}  e^{-ik_2x} 
 +   a^\dagger_{k_1} e^{ik_1x} e^{\frac{i}{2} (ik_1) \wedge (ik_2)}  b^\dagger_{k_2} e^{ik_2x} \right]
\label{proof11}
\end{eqnarray}
The operators $a_p$ and $b_p$ satisfy twisted commutation relations, so using (\ref{cstcom}) in (\ref{identity1}) we have 
\begin{eqnarray}
\phi^\dagger_\theta (x) \ast \phi_\theta (x) & = & \int d^3 \tilde{k}_1 d^3 \tilde{k}_2 \left[ a_{k_2}  b_{k_1}  e^{\frac{i}{2} (k_1) \wedge (k_2)}  e^{-ik_1x} \,e^{-ik_2x}
  +  b^\dagger_{k_2} \, b_{k_1} e^{\frac{-i}{2} (k_1) \wedge (k_2)} e^{-ik_1x} e^{ik_2x} \right. \nonumber \\
& - & (2\pi)^3 \, 2 E_{k_1} \, \delta^3 (k_1 - k_2)\, e^{-ik_1x} \,e^{ik_2x} \, + \,  a_{k_2} \, a^\dagger_{k_1}\, e^{\frac{-i}{2} (k_1) \wedge (ik_2)} \,e^{ik_1x} \, e^{-ik_2x}
\nonumber \\
 & + & \left. (2\pi)^3 \, 2 E_{k_1} \, \delta^3 (k_1 - k_2)\,e^{ik_1x} \, e^{-ik_2x} 
\, + \,   b^\dagger_{k_2}\,  a^\dagger_{k_1} \, e^{\frac{i}{2} (k_1) \wedge (k_2)}\, e^{ik_1x}\,e^{ik_2x}  \right] \nonumber \\
& = &  \int d^3 \tilde{k}_2 \left[ a_{k_2} \, e^{-ik_2x} \, + \, b^\dagger_{k_2} e^{ik_2x}  \right] \, e^{\frac{i}{2} \,\overleftarrow{\partial} \wedge \overrightarrow{\partial}}  \,
 \int d^3 \tilde{k}_1 \left[ b_{k_1} \, e^{-ik_1x} \, + \, a^\dagger_{k_1} e^{ik_1x} \right] \nonumber \\
& = &  \phi_\theta (x) \, \ast \, \phi^\dagger_\theta (x) 
\label{proof1}
\end{eqnarray}

\begin{eqnarray}
 \textbf{2)} \quad & \left [ \phi^\dagger_0 (x) \, \phi_0 (x) \right] \, e^{\frac{1}{2}\, \overleftarrow{\partial} \wedge P}
\,  = \, \left [ \phi_0 (x) \, \phi^\dagger_0 (x) \right] \, e^{\frac{1}{2}\, \overleftarrow{\partial} \wedge P} & \qquad \qquad \qquad 
\qquad \qquad \qquad 
\label{identity2}
\end{eqnarray}

One can check this identity by explicit calculations. But in view of (\ref{identity1}), this is easily checked to be true. Indeed this is nothing but (\ref{identity1}) rewritten 
in terms of commutative fields using dressing transformations (\ref{csfdress}). \\

These two identities can be generalized to a product of arbitrary number of fields. Hence for a string of fields we have 
\begin{eqnarray}
 \textbf{3)} \qquad \quad \phi^\dagger_\theta (x) \, \ast \, \phi_\theta \,  \dots \, \phi^\dagger_\theta (x) \, \ast \, \phi_\theta  
\, & = &\,  \phi^\dagger_\theta (x) \, \ast \, \phi_\theta \, \dots  \,\phi_\theta (x) \, \ast \, \phi^\dagger_\theta (x) \qquad \qquad \qquad \qquad  
\nonumber \\
\qquad & = & \phi_\theta (x) \, \ast \, \phi^\dagger_\theta (x) \, \dots \, \phi_\theta (x) \, \ast \, \phi^\dagger_\theta (x)  \nonumber \\
\qquad & = & \text{Other Permutations.}
\label{identity3}
\end{eqnarray}
Using dressing transformation (\ref{csfdress}),  (\ref{identity3}) can be rewritten in terms of the commutative fields, so that 
\begin{eqnarray}
 \textbf{4)} \qquad \quad \left [ \phi^\dagger_0 (x) \, \phi_0 (x) \, \dots \phi^\dagger_0 (x) \, \phi_0 (x) \right] \, e^{\frac{1}{2}\, \overleftarrow{\partial} \wedge P}
& = &  \left [ \phi^\dagger_0 (x) \, \phi_0 (x) \, \dots \phi_0 (x) \, \phi^\dagger_0 (x) \right] \, e^{\frac{1}{2}\, \overleftarrow{\partial} \wedge P} 
  \nonumber \\
& = &  \left [\phi_0 (x) \, \phi^\dagger_0 (x) \dots \phi^\dagger_0 (x) \, \phi_0 (x) \,\right] \, e^{\frac{1}{2}\, \overleftarrow{\partial} \wedge P} \nonumber \\
& = & \text{Other Permutations.} 
\label{identity4}
\end{eqnarray}

From (\ref{identity3}) it is clear that inspite of $\phi_\theta$ not satisfying usual commutation relation (\ref{op}), the six possible apparently different terms in 
(\ref{csnchamiltonian}) are one and the same. Hence (\ref{csnchamiltonian}) simplifies to
 \begin{eqnarray}
 \mathcal{H}^{\theta}_{\rm{Int}} & = & \left\{\frac{\lambda_1}{4} \, + \, \frac{\lambda_2}{4} \, + \, \frac{\lambda_3}{4}  \, + \, \frac{\lambda_4}{4}  \, + \, \frac{\lambda_5}{4}  
\, + \,\frac{\lambda_6}{4}\right\} \, \phi^\dagger_{\theta} \ast \phi^\dagger_{\theta} \ast \phi_{\theta} \ast \phi_{\theta} (x)  \nonumber \\
& = &  \frac{\lambda}{4} \, \phi^\dagger_{\theta} \ast \phi^\dagger_{\theta} \ast \phi_{\theta} \ast \phi_{\theta} (x),
\label{csncham}
\end{eqnarray}
where $ \lambda = \lambda_1 \, + \, \lambda_2 \, + \,\lambda_3 \, + \,\lambda_4 \, + \, \lambda_5 \, + \, \lambda_6 $.

One can further simplify (\ref{csncham}) using the dressing transformation, so that
\begin{eqnarray}
  \mathcal{H}^{\theta}_{\rm{Int}} & = & \frac{\lambda}{4} \,\phi^\dagger_{\theta} \ast \phi^\dagger_{\theta} \ast \phi_{\theta} \ast \phi_{\theta} (x) \nonumber \\
& = & \frac{\lambda}{4} \,\int d^3 \tilde{k}_{1} \left[ b_{k_{1}} \, e^{-ik_{1}x} \, + \, a^\dagger_{k_{1}} e^{ik_{1}x} \right] 
\,  e^{\frac{i}{2}\,\overleftarrow{\partial}\wedge\overrightarrow{\partial}} \left\{ \,
\int d^3 \tilde{k}_{2} \left[ b_{k_{2}} \, e^{-ik_{2}x} \, + \, a^\dagger_{k_{2}} e^{ik_{2}x} \right] \right.\nonumber \\
& &  \left.e^{\frac{i}{2} \overleftarrow{\partial}\wedge\overrightarrow{\partial}} \left\{
 \int d^3 \tilde{k}_3 \left[ a_{k_3}  e^{-ik_3x}  +  b^\dagger_{k_3} e^{ik_3x}  \right]  e^{\frac{i}{2}\,\overleftarrow{\partial}\wedge\overrightarrow{\partial}} 
\int d^3 \tilde{k}_4 \left[ a_{k_4}  e^{-ik_4x} \right.  \right.  \right. \nonumber \\
& & + \, \left. \left. \left. b^\dagger_{k_4} e^{ik_4x}  \right] \right\} \right\}.
\label{i1}
 \end{eqnarray}
Now, let us take a generic term like
 \begin{eqnarray}
 & &  b_{k_{1}} \, e^{-ik_{1}x} \, e^{\frac{i}{2} \,\overleftarrow{\partial} \wedge \overrightarrow{\partial}}  \, \left\{  a^\dagger_{k_{2}} \, e^{ik_{2}x} 
\,  e^{\frac{i}{2}\,\overleftarrow{\partial} \wedge \overrightarrow{\partial}}  \, \left\{ b^\dagger_{k_3} \, e^{ik_3x} 
\,  e^{\frac{i}{2}\,\overleftarrow{\partial} \wedge \overrightarrow{\partial}} \,  a_{k_4} \, e^{-ik_4x}  \right\} \right\} \nonumber \\
& = &  d_{k_{1}} \, e^{-\frac{i}{2} \, k_1 \wedge P}  \, e^{-ik_{1}x} \, e^{\frac{i}{2} \,\overleftarrow{\partial} \wedge \overrightarrow{\partial}}  \,
\left\{ c^\dagger_{k_{2}} \, e^{\frac{i}{2} \, k_2 \wedge P} \, e^{ik_{2}x} \, e^{\frac{i}{2} \,\overleftarrow{\partial} \wedge \overrightarrow{\partial}}  \,
\left\{ d^\dagger_{k_{3}} \, e^{\frac{i}{2} \, k_3 \wedge P} \, e^{ik_3x} \right. \right. \nonumber \\
& & \left. \left. e^{\frac{i}{2} \,\overleftarrow{\partial} \wedge \overrightarrow{\partial}}  \,  c_{k_{4}} \, e^{-\frac{i}{2} \, k_4 \wedge P} \, e^{-ik_4x} \right\} \right\}
\nonumber \\
& = &  d_{k_{1}} \, e^{-ik_{1}x} \, e^{-\frac{i}{2} \, k_1 \wedge P} \, e^{\frac{i}{2} \, (-ik_1) \wedge (ik_2 + ik_3 -ik_4)} 
\, c^\dagger_{k_{2}} \, e^{ik_{2}x} \, e^{\frac{i}{2} \, k_2 \wedge P} \, e^{\frac{i}{2} \, ( ik_2) \wedge ( ik_3 -ik_4)} \nonumber \\
& & d^\dagger_{k_{3}} \, e^{ik_3x}\, e^{\frac{i}{2} \, k_3 \wedge P} \, e^{\frac{i}{2} \, (ik_3) \wedge (-ik_4)}\, c_{k_{4}} \, e^{-ik_4x}\, e^{-\frac{i}{2}\,k_4 \wedge P}
 \nonumber \\
& = &  d_{k_{1}} \, e^{-\frac{i}{2} \, k_1 \wedge P}  \, c^\dagger_{k_{2}} \, e^{\frac{i}{2} \, k_2 \wedge P} \, d^\dagger_{k_{3}} \, e^{\frac{i}{2} \, k_3 \wedge P} \,
c_{k_{4}} \, e^{-\frac{i}{2} \, k_4 \wedge P} \, e^{-ik_{1}x} \, e^{ik_{2}x} \, e^{ik_3x}\, e^{-ik_4x}\, \nonumber \\
& &  e^{\frac{i}{2} \, (-ik_1) \wedge (ik_2 + ik_3 -ik_4)} \, e^{\frac{i}{2} \, ( ik_2) \wedge ( ik_3 -ik_4)}\, e^{\frac{i}{2} \, ( ik_3) \wedge (-ik_4)}
\label{1genterm}
 \end{eqnarray}
To simplify it further, we need the identities 
 \begin{eqnarray}
e^{\frac{i}{2} q \wedge P} \, c_p \, e^{\frac{-i}{2} q \wedge P} & = & e^{\frac{-i}{2} q \wedge p} \, c_p \nonumber \\
e^{\frac{i}{2} q \wedge P} \, d_p \, e^{\frac{-i}{2} q \wedge P} & = & e^{\frac{-i}{2} q \wedge p} \, c_p \nonumber \\
e^{\frac{i}{2} q \wedge P} \, c^\dagger_p \, e^{\frac{-i}{2} q \wedge P} & = & e^{\frac{i}{2} q \wedge p} \, c^\dagger_p \nonumber \\
e^{\frac{i}{2} q \wedge P} \, d^\dagger_p \, e^{\frac{-i}{2} q \wedge P} & = & e^{\frac{i}{2} q \wedge p} \, d^\dagger_p.
\label{ncidentities}
 \end{eqnarray} 
Using (\ref{ncidentities}) in (\ref{1genterm}) we obtain
\begin{eqnarray}
 & & b_{k_{1}} \, e^{-ik_{1}x} \, e^{\frac{i}{2} \,\overleftarrow{\partial} \wedge \overrightarrow{\partial}}  \, \left\{  a^\dagger_{k_{2}} \, e^{ik_{2}x} 
\,  e^{\frac{i}{2}\,\overleftarrow{\partial} \wedge \overrightarrow{\partial}}  \, \left\{ b^\dagger_{k_3} \, e^{ik_3x} 
\,  e^{\frac{i}{2}\,\overleftarrow{\partial} \wedge \overrightarrow{\partial}} \,  a_{k_4} \, e^{-ik_4x}  \right\} \right\} \nonumber \\
& = & d_{k_{1}} \, c^\dagger_{k_{2}} \, d^\dagger_{k_{3}} \, c_{k_{4}} \, e^{-ik_{1}x} \, e^{ik_{2}x} \, e^{ik_3x}\, e^{-ik_4x}\, e^{\frac{1}{2} (-ik_1 + ik_2 + ik_3 - ik_4) \wedge P}
\nonumber \\
& = & d_{k_{1}} \, c^\dagger_{k_{2}} \, d^\dagger_{k_{3}} \, c_{k_{4}} \, e^{-ik_{1}x} \, e^{ik_{2}x} \, e^{ik_3x}\, e^{-ik_4x}\, e^{\frac{1}{2} \overleftarrow{\partial} \wedge P}
\label{11genterm}
\end{eqnarray}
One can check by similar computations that each and every term in (\ref{i1}) can be similarly simplified. Hence for a generic string of creation/annihilation operators we have
\begin{eqnarray}
  (a_1)^{\#}_{k_1}\, (a_2)^\#_{k_2} \dots (a_4)^\#_{k_4} \, e^{i(\pm k_1 \pm k_2 \dots \pm k_4)x} 
 =   (c_1)^\#_{k_1}\, (c_2)^\#_{k_2} \dots (c_4)^\#_{k_4} 
\, e^{i(\pm k_1 \pm k_2 \dots \pm k_4)x} \, e^{\frac{1}{2} \overleftarrow{\partial} \wedge P}
\label{genncop}
\end{eqnarray}
where $a^\#$ represents any of the twisted creation/annihilation operators and $c^\#$ is the analogous commutative operator.

Therefore using (\ref{11genterm}) and its generalized form (\ref{genncop}), (\ref{i1}) can be simplified to
\begin{eqnarray}
 \mathcal{H}^{\theta}_{\rm{Int}} & = & \frac{\lambda}{4} \, \phi^\dagger_{\theta} \ast \left(\phi^\dagger_{\theta} \ast \left(\phi_{\theta} \ast \phi_{\theta} \right) \right) \nonumber \\
&= & \left[\frac{\lambda}{4} \, \phi^\dagger_{0} \phi^\dagger_{0} \phi_{0} \phi_{0} \right] \, e^{\frac{1}{2} \overleftarrow{\partial} \wedge P} \nonumber \\
& = & \mathcal{H}^{0}_{\rm{Int}}\, e^{\frac{1}{2} \overleftarrow{\partial} \wedge P}
\label{simp1l}
\end{eqnarray}
where $\mathcal{H}^{0}_{\rm{Int}}  =  \frac{\lambda}{4} \, \phi^\dagger_{0} \phi^\dagger_{0} \phi_{0} \phi_{0} $ is the analogous commutative hamiltonian density.


\subsection*{The S-matrix}

The computation of S-matrix in this case is quite similar to that of real scalar fields discussed in earlier sections.

For a process of two-to-two particle scattering, the S-matrix elements are given by  
\begin{eqnarray}
 S_{\theta}[p_{2}, p_{1} \rightarrow  p'_{1}, p'_{2}] & \equiv &  S_{\theta}[p'_{2}, p'_{1} ; p_{2}, p_{1}]
\; = \;  \leftidx{_{\rm{out},\theta}}{\left \langle  p'_{2}, p'_{1} | p_{2}, p_{1} \right \rangle }{_{ \theta,\rm{in}}}
\label{csncsm}
\end{eqnarray}
where $ |p'_{1}, p'_{2} \rangle_{\theta,\rm{out}} $ is the noncommutative two particle out-state which is measured in the far future and $ |p_{2}, p_{1}  \rangle_{\theta,\rm{in}} $ 
is the noncommutative two particle in-state prepared in the far past.

Just like the case of real scalar fields, the noncommutative in- and out-states can be related with each other using S-matrix $\hat{S}_{\theta}$. Therefore we have
\begin{eqnarray}
 S_{\theta}[p'_{2}, p'_{1} ; p_{2}, p_{1}] & = &  \leftidx{_{\rm{out},\theta}}{\left \langle  p'_{2}, p'_{1} | \hat{S}_{\theta} | p_{2}, p_{1}  \right \rangle}{_{\rm{out},\theta}}
\; = \;  \leftidx{_{\rm{in},\theta}}{\left \langle  p'_{2}, p'_{1} | \hat{S}_{\theta} | p_{2}, p_{1}  \right \rangle}{_{\rm{in},\theta}}
\label{csncsmat}
\end{eqnarray}
where the noncommutative S-matrix $\hat{S}_{\theta}$, in interaction picture, can be written as 
\begin{eqnarray}
 \hat{S}_{\theta} & = & \lim_{t_{1} \rightarrow \infty} \lim_{t_{2} \rightarrow -\infty} U_{\theta}(t_{1},t_{2}) \nonumber \\
& = & \mathcal{T}  \exp \left[-i\int^{\infty}_{-\infty} d^{4}z \mathcal{H}^{\theta}_{\rm{Int}} (z) \right] \nonumber \\
& = & \mathcal{T}  \exp \left[ -i\int^{\infty}_{-\infty} d^{4}z \mathcal{H}^{0}_{\rm{Int}} (z)\, e^{\frac{1}{2} \overleftarrow{\partial_{z}} \wedge P }\right] 
\label{csncsint}
\end{eqnarray} 
where $\mathcal{H}^{\theta}_{\rm{Int}}$ is given by (\ref{simp1l}) and $\mathcal{H}^{0}_{\rm{Int}} (z)$ is its commutative analogue.

We can formally expand the exponential and write the $\hat{S}$ as a time-ordered power series given by
\begin{eqnarray}
 \hat{S}_{\theta} & = &   \mathbb{I} \, + \, -i\int^{\infty}_{-\infty} d^{4}z  \mathcal{H}^{0}_{\rm{Int}} (z)\, e^{\frac{1}{2} \overleftarrow{\partial_{z}} \wedge P } 
\nonumber \\
& + &  \mathcal{T}\, (-i )^2 \int^{\infty}_{-\infty} d^{4}z \int^{\infty}_{-\infty} d^{4}z'\, \mathcal{H}^{0}_{\rm{Int}} (z)\, e^{\frac{1}{2} \overleftarrow{\partial_{z}} \wedge P } 
\,  \mathcal{H}^{0}_{\rm{Int}} (z')\, e^{\frac{1}{2} \overleftarrow{\partial_{z'}} \wedge P } \, + \, \dots
\label{csspower}
\end{eqnarray}
Now let us take the second term and simplify it to 
\begin{eqnarray}
 -i\int^{\infty}_{-\infty} d^{4}z  \mathcal{H}^{0}_{\rm{Int}} (z)\, e^{\frac{1}{2} \overleftarrow{\partial_{z}} \wedge P } 
& = &   -i\int^{\infty}_{-\infty} d^{4}z  \mathcal{H}^{0}_{\rm{Int}} (z) 
\label{cssecpower}
\end{eqnarray}
where as done in \cite{bal-uvir} we have expanded the exponential, integrated and discarded all the surface terms. With computations analogous to that done in \cite{bal-uvir} one 
can similarly show that all the the higher order terms in power series of (\ref{csspower}) will be free of any $\theta$ dependence. We refer to \cite{bal-uvir} for more details. 

Hence we have
\begin{eqnarray}
 \hat{S}_{\theta} & = & \mathcal{T} \exp\left[-i\int^{\infty}_{-\infty} d^{4}z  \mathcal{H}^{0}_{\rm{Int}} (z)  e^{\frac{1}{2} \overleftarrow{\partial_{z}} \wedge P } \right]
  \; = \; \hat{S}_{0}
\label{csncsoperator}
\end{eqnarray}

Like the previously discussed real scalar field case,  here also the $\hat{S}_{\theta} $ turns out to be completely equivalent to $\hat{S}_{0}$.  
The noncommutative S-matrix elements have only overall noncommutative phases in them. This implies that there is no UV/IR mixing and the physical observables e.g 
scattering cross-section and decay rates are independent of $\theta$. 


\subsection{Yukawa Interactions}

Like the scalar fields, a noncommutative spinor field $\psi_\theta$ is composed of twisted fermionic creation/annihilation operators and has a normal mode expansion 
\begin{eqnarray}
 \psi_\theta (x) & = & \int d^3 \tilde{k} \sum_s \left[ a_{s,k} \, u_{s,k}\, e^{-ikx} \, + \, b^\dagger_{s,k} \, v_{s.k}\,e^{ikx} \right], \nonumber \\
\overline{\psi}_\theta (x) & = & \int d^3 \tilde{k} \sum_s \left[ b_{s,k} \, \overline{v}_{s,k}\, e^{-ikx} \, + \, a^\dagger_{s,k} \,\overline{u}_{s.k}\,e^{ikx} \right]
\label{ncspinor}
\end{eqnarray}
where  $u_{s,k}$ and $v_{s,k}$ are four component spinors (same as commutative case), $ d^3 \tilde{k} \, = \, \frac{d^3 k}{(2\pi)^3 2 E_k}$ and $a_{s,p}$, $b_{s,p}$ are twisted 
fermionic operators satisfying relations similar to (\ref{cstcom}) but with $\eta = - 1$. 

The operators $a_{s,p}$, $b_{s,p}$ can again be related with their commutative counterparts $c_{s,p}$, $d_{s,p}$ by dressing transformations similar to (\ref{csfdress}). Hence, 
$\psi_\theta$ can also be related with the commutative spinor field $\psi_0$ by
\begin{eqnarray}
 \psi_{\theta} (x) & = & \psi_{0} (x) \, e^{\frac{1}{2} \, \overleftarrow{\partial} \wedge P}, \nonumber \\
 \overline{\psi}_{\theta} (x) & = & \overline{\psi}_{0} (x) \, e^{\frac{1}{2} \, \overleftarrow{\partial} \wedge P}
\label{spfdress}
\end{eqnarray}
Using $\psi_{\theta}$ and $\phi_{\theta}$ we can construct a Yukawa interaction term given by
\begin{eqnarray}
  \mathcal{H}^{\theta}_{Yuk} & = & \eta_1 \, \overline{\psi}_{\theta} \ast \phi_{\theta} \ast \psi_{\theta}
\, + \, \eta_2 \, \overline{\psi}_{\theta} \ast \psi_{\theta} \ast \phi_{\theta} 
\label{yukawa}
\end{eqnarray}
 Using identities similar to (\ref{identity1}) - (\ref{identity4}) one can show that the two terms in (\ref{yukawa}) are the same, so that
\begin{eqnarray}
  \mathcal{H}^{\theta}_{Yuk} & = & \eta \, \overline{\psi}_{\theta} \ast \phi_{\theta} \ast \psi_{\theta}
\label{yukawa1}
\end{eqnarray}
with $\eta = \eta_1 + \eta_2$. Using the dressing transformation (\ref{spfdress}) one can see that
\begin{eqnarray}
\mathcal{H}^{\theta}_{Yuk} & = & \left[ \eta \, \overline{\psi}_{0} \phi_{0} \psi_{0} \right] \, e^{\frac{1}{2} \, \overleftarrow{\partial} \wedge P} 
\; = \; \mathcal{H}^{0}_{Yuk} \, e^{\frac{1}{2} \, \overleftarrow{\partial} \wedge P}
\label{yukawa2}  
\end{eqnarray}
where $\mathcal{H}^{0}_{Yuk} = \eta \, \overline{\psi}_{0} \phi_{0} \psi_{0} $ is the commutative Yukawa interaction term.

We again find that the noncommutative interaction hamiltonian density is (analogous commutative hamiltonian density) $\times$ $e^{\frac{1}{2} \overleftarrow{\partial} \wedge P}$. 
By computations similar to that done before we have 

\begin{eqnarray}
 \hat{S}_{\theta} & = & \mathcal{T}  \exp \left[-i\int^{\infty}_{-\infty} d^{4}z \mathcal{H}^{\theta}_{Yuk} (z) \right] \nonumber \\
& = & \mathcal{T}  \exp \left[ -i\int^{\infty}_{-\infty} d^{4}z  \mathcal{H}^{0}_{Yuk} (z)\, e^{\frac{1}{2} \overleftarrow{\partial_{z}} \wedge P }\right] \nonumber \\
& = &  \hat{S}_{0}
\label{yuksop}
\end{eqnarray} 
Since $\hat{S}_{\theta} = \hat{S}_{0}$, the S-matrix elements for any process have only overall noncommutative phases coming due to the twisted statistics 
of the in- and out-states. 

The equivalence between $\hat{S}_{\theta}$ and $\hat{S}_{0}$ and the fact that only an overall noncommutative phase appears in S-matrix elements is a generic result. It holds true 
for any noncommutative field theory having polynomial interactions and involving only matter fields \cite{ssb}.

\subsection{Renormalization}

Since for any pure matter theory having polynomial interactions, the $\hat{S}_{\theta}$ always turns out to be the same as $\hat{S}_{0}$ and the noncommutative S-matrix elements have only overall noncommutative 
phase dependences, the noncommutative 1PI functions also have only overall noncommutative phase dependences. Apart from the divergences already present in analogous commutative 
theories, there are no new source of divergences, in any such noncommutative theory. So all such theories are renormalizable provided the analogous 
commutative theory is itself renormalizable. Moreover, as we saw from explicit calculations for the case of $\phi^4_{\theta,\ast}$ theory, the essential techniques of 
renormalization remains the same as the commutative ones and these noncommutative theories can always be renormalized in a way very similar to the commutative theories.

Also, as in case of $\phi^4_{\theta,\ast}$ theory, the $\theta$ dependent phases present in 1PI functions for all such theories will sit more like spectators and will not 
change the $\beta$-functions, RG flow of couplings or the fixed points, from those of the analogous commutative theory.


\section{Conclusions}

In this work we have presented a complete and comprehensive treatment of noncommutative theories involving only matter fields. We have shown first for real scalar fields having a 
$\phi^4_{\theta,\ast}$ interaction and then for more generic theories that the noncommutative $\hat{S}_{\theta}$ is the same as $\hat{S}_{0}$ and that the S-matrix elements only have 
an overall phase dependence on the noncommutativity scale $\theta$. We have also argued that since there is only an overall phase dependence on the noncommutativity scale $\theta$, 
there are no non-planar diagrams and hence complete absence of UV/IR mixing in any such theory. 

We have further showed that all such theories are renormalizable if and only if the corresponding commutative theories are renormalizable. The usual commutative techniques for 
renormalization can be used to renormalize such theories. Moreover, we showed by explicit calculations for $\phi^4_{\theta,\ast}$ case and argued for generic case, that for 
all such theories the $\beta$-functions, RG flow of couplings or the fixed points are all same as those of the analogous commutative theory. 

It should be further remarked that since the noncommutative S-matrix elements differ from the analogous commutative ones by only an overall $\theta$ dependent phase,
hence some observables like transition probabilities, scattering cross-sections, decay rates etc remain same. 
The equivalence of the above physical observables along with that of $\beta$-functions, RG flow and fixed points with those of the corresponding 
commutative theories does not mean that the all such noncommutative theories are one and the same as their commutative counterparts. 
There still exist various other observables in the theory which differ in the noncommutative case from the commutative ones. For example, the appearance of only overall phase factors  
in S-matrix elements is due to the fact that we chose to work with definite momentum states. If we had taken 
wavepackets instead of plane wave states, then we would not have been able to pull out an overall noncommutative phase factor and we would have obtained nontrivial dependence
on the noncommutative $\theta$ parameters. Since in this work we were interested in studying the renormalization of twisted theories and not in looking for phenomenological 
signatures of noncommutativity, so we chose to work with plane wave states instead of wavepackets to avoid unnecessary complications in our investigations. Moreover, even if
one chooses to work with plane waves, the resulting overall noncommutative phases in the S-matrix elements will result in change in the time delay in decay processes.

Also, one can always construct, even for free theories, appropriate observables, which unambiguously distinguish between a noncommutative and commutative theory. For instance, 
the twisted statistics of the particles will result in violation in Pauli principle \cite{bal,pramod}, changes in HBT correlations \cite{rahul} as well as changes in various other thermodynamic quantities \cite{basu,basu1}.
The noncommutativity is also expected to have nontrivial signatures in CMB spectrum \cite{anosh} etc. Moreover, as shown in \cite{amilcar-pinzul} if one considers nonabelian gauge theories coupled with matter fields, then, 
indeed, there are nontrivial dynamical dependences on $\theta$ parameters. 

The discussion of this paper was limited only to matter fields and polynomial interaction terms constructed out of only matter fields. Noncommutative field theories involving nonabelian gauge 
fields violate twisted Poincar\'e invariance and are know to suffer from UV/IR mixing \cite{amilcar-pinzul}. They require special treatment which is outside the scope of present 
work. We plan to discuss gauge theories in a future work.


\section*{Acknowledgements}

It is a pleasure to thank A. P. Balachandran, T. R. Govindarajan and P. Padmanabhan for many useful discussions and critical comments. RS will also like to thank 
M. Paranjape and Groupe de Physique des Particules of Universit\'e de Montr\'eal, where a part of the work was done, for their wonderful hospitality. SV would like to thank 
ICTP, Trieste for support during the final stages of this work.
ARQ is supported by  CNPq under grant number 307760/2009-0. RS is supported by C.S.I.R under the award no: F. No 10 – 2(5)/2007(1) E.U. II.  \\ 



\end{document}